\documentclass[fleqn,usenatbib]{mnras}

\usepackage{newtxtext,newtxmath}
\usepackage[T1]{fontenc}

\DeclareRobustCommand{\VAN}[3]{#2}
\let\VANthebibliography\thebibliography
\def\thebibliography{\DeclareRobustCommand{\VAN}[3]{##3}\VANthebibliography}

\usepackage{graphicx}	% Including figure files
\usepackage{amsmath}	% Advanced maths commands
\newcommand{\st}{{\it St}}
\title{On the non-axisymmetric fragmentation of rings generated by the  Secular Gravitational Instability}
\author[A. Pierens]{Arnaud Pierens $^{1}$\thanks{E-mail:arnaud.pierens@u-bordeaux.fr} \\
$^{1}$Universit\'e de Bordeaux, Observatoire Aquitain des Sciences de l'Univers,
    All\'ee Geoffroy St. Hilaire, 33165 Pessac, France}
\pagerange{\pageref{firstpage}--\pageref{lastpage}} 

\def\LaTeX{L\kern-.36em\raise.3ex\hbox{a}\kern-.15em 
    T\kern-.1667em\lower.7ex\hbox{E}\kern-.125emX}
\begin{document}
\label{firstpage}
\maketitle
\begin{abstract}
Ringed structures have been observed in a variety of protoplanetary discs. Among the processes that might be able to generate such features, the Secular Gravitational Instability (SGI) is a possible candidate. It has also been proposed that the SGI might lead to the formation of planetesimals during the non-linear phase of the instability. In this context, we employ two-fluid hydrodynamical simulations with self-gravity to study the non-axisymmetric, non-linear evolution of ringed perturbations that grow under the action of the SGI.  We find that the non-linear evolution outcome of the SGI depends mainly on the initial linear growth rate. For SGI growth rates smaller than typically $\sigma \gtrsim 10^{-4}-10^{-5}\Omega$, dissipation resulting from dust feedback introduces a $m=1$ spiral wave in the gas, even for Toomre gas stability parameters $Q_g>2$ for which non-axisymmetric instabilities appear in a purely gaseous disc. This one-armed spiral subsequently traps dust particles until a dust-to-gas ratio $\epsilon \sim 1$ is achieved.   For higher linear growth rates, the dust ring  is found to undergo gravitational collapse until the  bump in the surface density profile becomes strong enough to trigger the formation of dusty vortices through the Rossby Wave Instability (RWI).  Enhancements in dust density resulting from this process are found to  scale with the linear growth rate, and can be such that the dust density is higher than the Roche density, leading to the formation of  bound clumps.  Fragmentation of axisymmetric rings produced by the  SGI might therefore appear as a possible process for the formation of planetesimals.
\end{abstract}
\begin{keywords}
accretion, accretion discs --
                planet-disc interactions--
                planets and satellites: formation --
                hydrodynamics --
                methods: numerical
\end{keywords}

\section{Introduction}

In the standard scenario for planet formation, mm-sized dust grows  from the coagulation of micron-sized grains (Dullemond \& Dominik 2005).  However, further growth  of particles in the mm-size range is limited because of the bouncing  (Zsom et al. 2010) and fragmentation (Blum \& Wurm 2008) barriers.   An emerging picture to bypass these growth barriers is that 100-km sized planetesimals form directly through the streaming instability (Youdin \& Goodman 2005; Johansen et al. 2009; Simon et al. 2016) in which particles with Stokes number (or dimensionless stopping time) $\st\sim 0.001-0.1$  directly concentrate into clumps or filaments under the action of gas drag  and which can subsequently  become gravitationally unstable  to form $\sim100-1000$ km size bodies. These planetesimals can subsequently grow very efficiently by capturing inward drifting pebbles (Johansen \& Lacerda 2010; Lambrechts \& Johansen 2012), namely solids with Stokes number $\st\sim 0.01-1$ that are marginally coupled to the gas,  leading eventually to the formation of  giant planet cores within 1 Myr (Lambrechts \& Johansen 2014). \\

The conditions for triggering the SI, and the efficiency of pebble accretion, are however sensitive to the level of turbulence operating in the disc. Small particles are lofted away from the midplane by turbulent mixing, such that the local dust-to-gas ratio, and hence the growth rate of the SI, are reduced. One possibility to counteract the effect of turbulence in the disc is to invoke the presence of pressure bumps in the disc, where the radial drift of dust particles can be stopped, leading subsequently to an enhancement of dust-to-gas ratio.

Such pressure bumps could be associated with ringed structures,  which have been observed with ALMA in a variety of protoplanetary discs (ALMA Partnership et al. 2015; Andrews et al. 2016,2018). Although there is so far no adopted consensus, possible mechanisms that might be able to explain these features include: the presence of planets that can open gaps in the disc (Dipierro et al. 2015; Dong et al. 2017), zonal flows (Flock et al. 2015), snow lines (Zhang et al. 2015), large-scale instabilities (L\`oren-Aguilar \& Bate 2016), 
spontaneous ring formation through radial drift plus dust coagulation (Gonzalez et al. 2017), dust-driven viscous ring-instability (Dullemond \& Penzlin 2018). \\

The Secular Gravitational Instability (hereafter SGI), which is a two-fluid instability and occurs because of  self-gravity and dissipation from gas drag (Takahashi \& Inutsuka 2014, 2016) is an alternative process to induce ringed structures in protoplanetary discs. The linear growth stage of SGI has been widely examined using both one-fluid (Youdin 2011; Shariff \& Cuzzi 2011; Michikoshi et al. 2012) and two-fluid models (Takahashi \& Inutsuka 2014; Latter \& Rosca 2017). In particular, results of two-fluid models suggest that the SGI can be a good candidate in producing dusty rings at distances $\gtrsim 10$ AU  provided that the turbulence operating  in the disc is relatively weak, with $\alpha$ viscous stress parameter (Shakura \& Sunyaev 1973) $\lesssim 10^{-5}$ typically. The axisymmetric, non-linear evolution of the SGI was examined by Tominaga et al. (2018) who found that the dust tends to collapse during the non-linear stage, leading to an increase of the dust density by a factor of $\sim 100$  in discs with zero pressure gradient. In a following study,  Tominaga et al. (2020) has confirmed that the  SGI can still create dusty rings in more realistic models where radial drift of solids is taken into account,  although the dust concentration enhancement is somewhat smaller in that case. \\

In this short paper, we examine the non-linear evolution of the SGI in the non-axisymmetric case. Non-axisymmetric linear analysis of the SGI  was studied by Shadmehri et al. (2019) who found that non-axisymmetric perturbations  can exhibit a significant growth during a finite time interval even when the system is stable against axisymmetric instability. Here our aim is rather to check, as speculated by Latter \& Rosca (2017),  whether or not axisymmetric rings produced by the SGI break into clumps during the non-linear phase, leading to the possible formation of planetesimals. We find that it is indeed the case and that dusty clumps can be easily formed once the amplitude of the axisymmetric ring perturbation becomes high enough to  trigger the  Rossby Wave Instability (RWI; Lovelace et al. 1999; Li et al. 2000). 

The paper is organized as follows.  In Sect.2, we present the governing evolution equations of the gas and dust components. We then describe in Sect. 3 the numerical setup and the initial conditions that are used in the simulations, whose results are presented in Sect. 4.  We discuss and draw our conclusions in Sect.5.

\section{Gas and dust disc models}
\label{sec:dustdisc}

We consider  razor-thin, locally isothermal protoplanetary disc models and make use of a two-fluid approach to follow the evolution of the gas and solid particles. For the particles, the equations for the conservation of mass and momentum are given by:

\begin{equation}
\frac{\partial \Sigma_p}{\partial t}+\nabla\cdot(\Sigma_p{\bf V})=\nabla\cdot\left[D\Sigma_g\nabla\left(\frac{\Sigma_p}{\Sigma_g}\right)\right]
\end{equation}
\begin{equation}
\frac{\partial {\bf V}}{\partial t}+({\bf V}\cdot \nabla){\bf V}=-\frac{\nabla P_p}{\Sigma_p}-\frac{{\bf V}-{\bf v}}{\tau_s}-{\bf \nabla} \Phi_{\star}-{\bf \nabla} \Phi_{sg}-{\bf \nabla} \Phi_{ind}
\label{eq:dust}
\end{equation}
where $\Sigma_p$ is the particle surface density, ${\bf V}$ the particle velocity, $P_p$ the particle pressure that we will write as $P_p=c_p^2\Sigma_p$, with $c_p$ the  particle velocity dispersion, $D$ the dust diffusion coefficient and $\tau_s$ the particle stopping time that we will parametrize through the Stokes number ${\it St}=\Omega_k \tau_s$ with 
$\Omega_k$ the Keplerian frequency.   The implication is  that we consider that the dust-gas coupling is fixed.  In a more realistic case of Epstein drag with fixed particle size, however, the stopping time should be proportional to the gas density, so that particles will become loosely coupled if they eventually fall in a gas-depleted region. 

Employing an $\alpha$ prescription (Shakura \& Sunyaev 1973) for the gas turbulent diffusion coefficient $D_g=\alpha c_s^2\Omega$, where $c_s$ is the gas sound speed, $c_p$ can be re-written in terms of $c_s$ and $\tau_s$ as (Youdin 2011; Latter \& Rosca 2017):
\begin{equation}
c_p=\frac{\sqrt{1+2\st^2+(5/4)\st^3}}{1+\st^2}\sqrt \alpha c_s
\label{eq:cp}
\end{equation}

while the dust diffusion coefficient is given by:

\begin{equation}
D=\frac{1+\st+4\st^2}{(1+\st^2)^2} D_g
\label{eq:d}
\end{equation}

Finally, in Eq. \ref{eq:dust},  $\Sigma_g$ and ${\bf v}$ refer  to the surface density and velocity of the gas respectively, and whose governing evolution equations are given by:

\begin{equation}
\frac{\partial \Sigma_g}{\partial t}+\nabla\cdot(\Sigma_g{\bf v})=0
\end{equation}
\begin{equation}
\frac{\partial {\bf v}}{\partial t}+({\bf v}\cdot \nabla){\bf v}=-\frac{\nabla P_g}{\Sigma_g}-\frac{\Sigma_p}{\Sigma_g}\frac{{\bf v}-{\bf V}}{\tau_s}
-{\bf \nabla} \Phi_{\star}-{\bf \nabla} \Phi_{sg}-{\bf \nabla} \Phi_{ind}
\end{equation}
where  $P_g=c_s^2 \Sigma_g$  the gas pressure.\\

 In the previous equations, $\Phi_\star$ is the gravitational potential due to the central star, $\Phi_{ind}$ is the indirect potential that results from the fact that the frame centred on the central star is not inertial and $\Phi_{sg}$ the self-gravitating potential which includes the contributions of both the gas and dust components. It is related to the dust and gas components through  the Poisson equation:

\begin{equation}
\nabla^2 \Phi_{sg}=4\pi G(\Sigma+\Sigma_p)
\label{eq:poisson}
\end{equation}

 \begin{figure*}
\centering
\includegraphics[width=\textwidth]{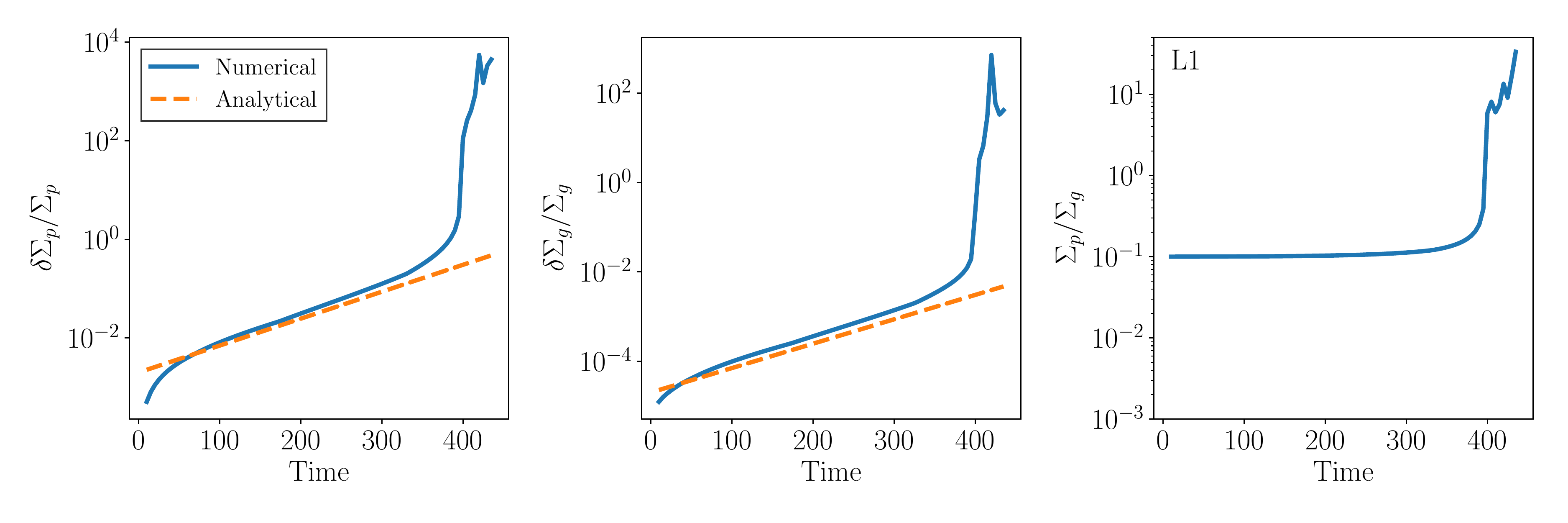}
\includegraphics[width=\textwidth]{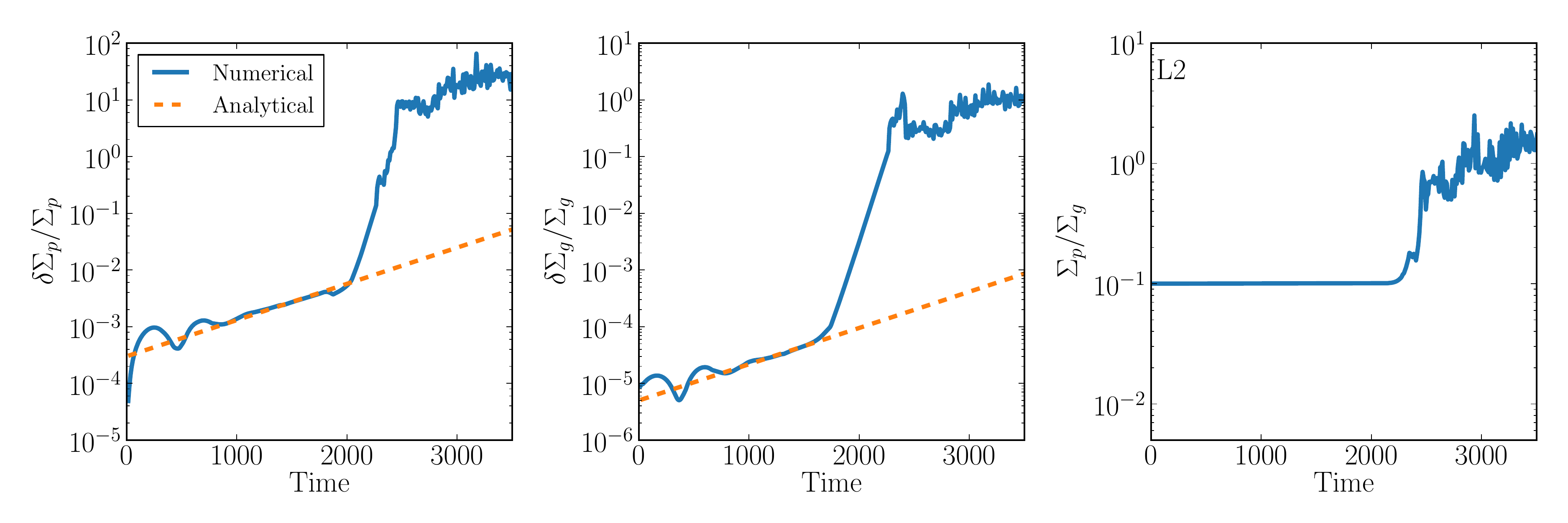}
\caption{{\it Top:} from left to right, temporal evolution of the dust surface density, gas surface density, and dust-to-gas ratio for model L1. {\it Bottom:} same but for model L2.}
\label{fig:linear}
\end{figure*}

\begin{figure}
\centering
\includegraphics[width=\columnwidth]{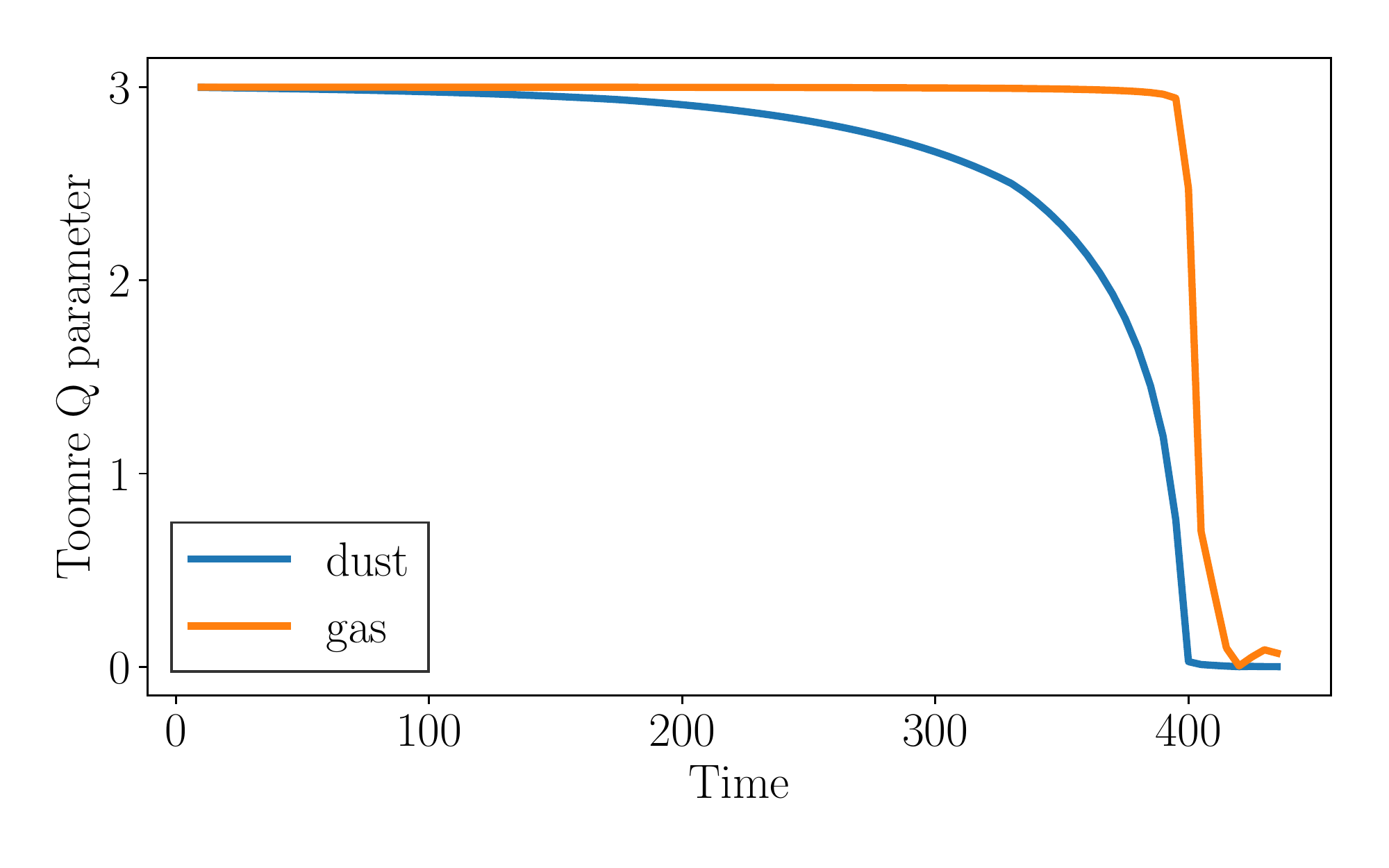}
\caption{For Model L1, temporal evolution of the Toomre stability parameter for  the dust and gas components.  }
\label{fig:toomre_b}
\end{figure}

\section{Numerical simulations}
\subsection{Numerical method}
Simulations were performed using the GENESIS (De Val-Borro et al. 2006) code which solves
the equations governing the disc evolution on a polar grid $(R,\varphi)$ using an advection scheme based on the monotonic  transport algorithm (Van Leer 1977). It uses the FARGO algorithm (Masset 2000) to avoid time step limitation due to the Keplerian velocity at the inner edge of the disc, and was recently extended to follow the evolution of a solid component that is modelled assuming a pressureless fluid.  Momentum exchange between the particles and the gas is handled by employing the semi-analytical scheme presented in Stoyanovskaya et al. (2018). This approach enables considering arbitrary solid concentrations and values for the Stokes number, and  is therefore very well suited for looking for solutions of non-stationary problems.  Tests of the numerical method to handle the momentum transfer between gas and dust have been presented in 
Pierens et al. (2019). \\

 The self-gravitational potential $\Phi_{sg}$ that includes the contributions of the two components (see Eq. \ref{eq:poisson}) is calculated using a Fast Fourier Transform (FFT) method (Binney \& Tremaine 1987; Baruteau \& Masset 2008). In order to take into account the effect of the finite disc thickness, the gravitational potential  is smoothed out using a softening length $r_{sm}=0.6H_p$ (Muller \& Kley 2012), where $H_P$ is the particle disc scale height. It is important to note that the same value should be chosen for the gas and solid smoothing lengths,  since in our work $\Sigma_g$ should be interpreted as the gas surface density located within the vertical extend 
 of the particle subdisc (Latter \& Rosca 2017). The contribution to the gravitational potential from the gas material located above and below the particle scale height is therefore not taken into account and considered as negligible. \\
 
 As required by the FFT method of Binney \& Tremaine (1987), a  logarithmic radial spacing is employed with a numerical domain extending between $R_{in} = 0.5$ and $R_{out} = 1.5$. Regarding the resolution, it is chosen such that the most unstable wavelength of the SGI is resolved by at least $\sim 64$ radial grid cells, {except for model R5 (see Sect. \ref{sec:concentration}) for which it is resolved by $\sim 50$ grid cells}. Test simulations indeed suggested that using such a resolution provides growth timescales that are in good agreement with analytical theory.\\
 
  The computational units that we adopt are such that the unit of mass is the central mass $M_\star$, the unit of distance is $R=1$ and the gravitational constant is $G=1$.  To present the results of
simulations we use the  orbital period  at $R_0=1$ as the unit of time. \\

 Regarding the boundary conditions, we make use of damping zones near the boundaries where the radial velocity and density for both the gas and particles are relaxed to their initial value. The buffer region extends from $R=0.50$ to $R=0.625$ at the inner boundary while it is located in the range $R\in [1.4,1.5]$ at the outer boundary. 
 
\begin{figure*}
\centering
\includegraphics[width=\textwidth]{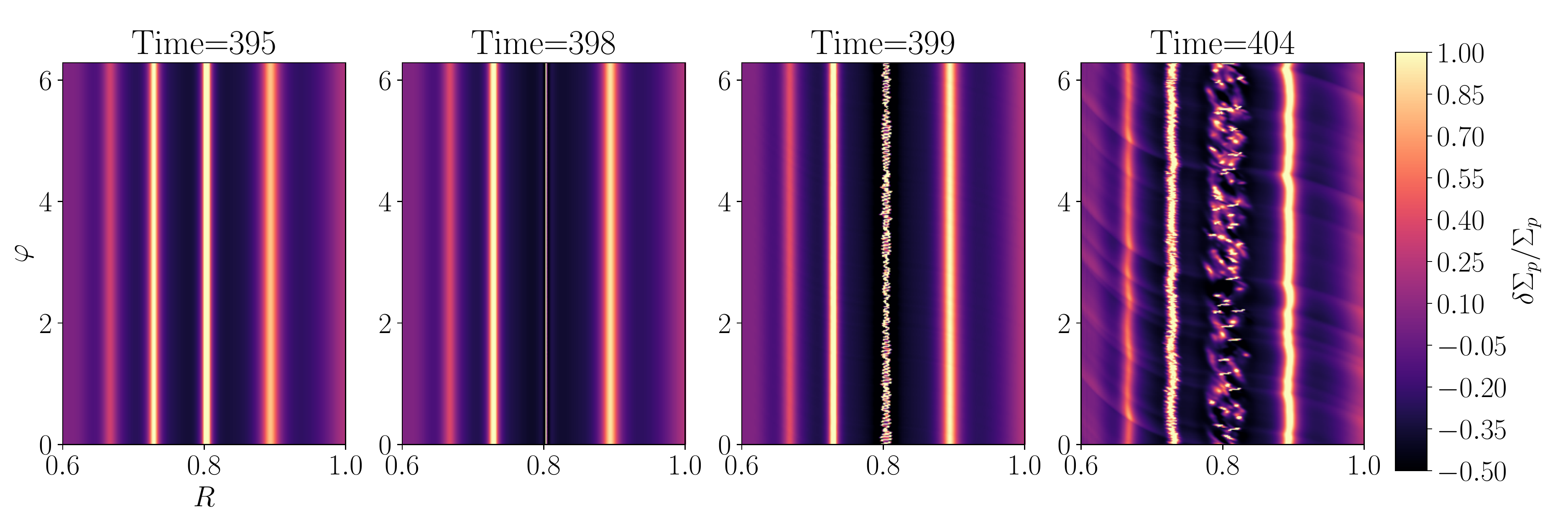}
\includegraphics[width=\textwidth]{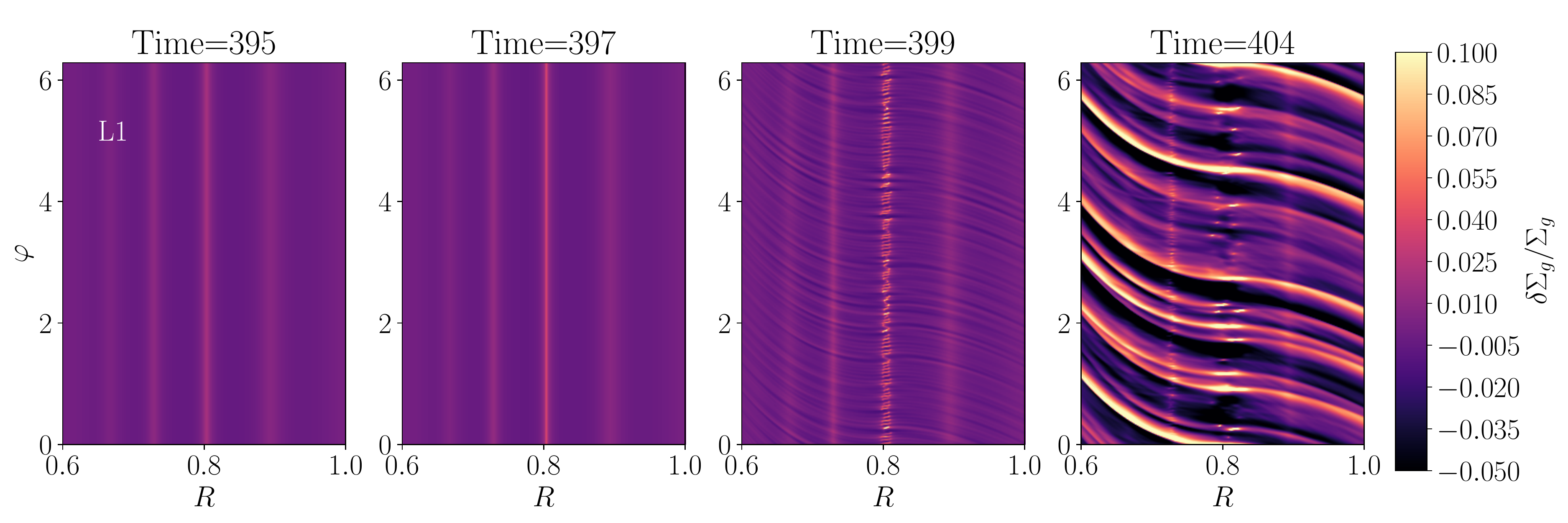}
\caption{{\it Top:} Relative dust density perturbation at different times for model L1.  {\it Bottom:} same but for the gas component.}
\label{fig:2d_modeleb}
\end{figure*}
 
 \subsection{Initial conditions}
 \label{sec:init}

 \indent \par{\bf Gas component--} We consider  locally isothermal disc models with aspect ratio  $h$ that varies as $h=0.07\left(\frac{R}{R_0}\right)^f$, where $f$ is the 
 disc flaring index which is set to $f=1$. The initial gas surface density is   $\Sigma_g=  \Sigma_{g,0}(R/R_0)^s$ , where $\Sigma_{g,0}$ is the initial surface density at  $R=R_0$ and where the power-law index is set to  $s=-1$.   The adopted gas disc model has therefore a  Toomre parameter   ${Q_g}=\Omega c_s/\pi G \Sigma_g$ which is constant across the numerical domain.

\par{\bf Solid component--} We consider particles with Stokes number in the range $\st \in[0.01,10]$. The initial particle surface density is such that the initial solid-to-gas ratio (or metallicity) $\epsilon=\Sigma_p/ \Sigma_g$  is constant throughout the disc, with $\epsilon\in [0.01,0.1]$. \\

For a given set  of input parameters $(\st, \epsilon)$, velocities for both the gas and solid components  can  then be initiated using the steady-state radial drift solution resulting from the coupling between the gas and solid component (Benitez-llambay et al. 2016; Pierens et al. 2019). These are given by (see also Kanagawa et al. 2017; Dipierro et al. 2018):

\begin{equation}
V_R=\frac{{\it St}}{(1+\epsilon^2)+{\it St}^2}\Delta v
%+\frac{1+\epsilon}{(1+\epsilon)^2+\st^2}v_{visc}
\label{eq:vrd}
\end{equation}
\begin{equation}
%V_\phi=\frac{1}{2}\left[\frac{1+\epsilon}{((1+\epsilon)^2+\st^2)}\Delta v
V_\varphi=\frac{1+\epsilon}{2((1+\epsilon)^2+\st^2)}\Delta v
% -\frac{v_{visc}}{(1+\epsilon)^2+\st^2}\right]
\end{equation}

for the particles and:
\begin{equation}
 v_R=-\frac{\epsilon \st}{(1+\epsilon)^2+\st^2}\Delta v
 %+\frac{v_{visc}}{1+\epsilon}\left(1+\epsilon\frac{\st^2}{(1+\epsilon)^2+\st^2}\right)
 \label{eq:vrg}
\end{equation}
\begin{equation}
%v_\phi=\frac{1}{2}\left[\frac{1}{(1+\epsilon)}\left(1+\frac{\epsilon \st^2}{(1+\epsilon)^2+\st^2}\right)\Delta v+\frac{\epsilon \st \; v_{visc}}{(1+\epsilon)^2+\st^2}\right)
v_\varphi=\frac{1}{2(1+\epsilon)}\left(1+\frac{\epsilon \st^2}{(1+\epsilon)^2+\st^2}\right)\Delta v
\label{eq:vtg}
\end{equation}

for the gas, with: 

\begin{equation}
\Delta v=\frac{1}{\Sigma_g\Omega}\frac{\partial P}{\partial r}=h^2v_k(2f+s-1)=-\eta v_k
\end{equation}

where $\Omega$ is the keplerian frequency. We note in passing that for the disc model that we adopt, namely for $s=-1$ and  $f=1$, $\Delta v=0$ such that dust does not drift initially.

\section{Results}

\subsection{Linear growth of the SGI}

\begin{table*}
\caption{Summary of parameters adopted in our runs. }              % title of Table
\label{table1}      % is used to refer this table in the text
\centering
%\resizebox{0.7\textwidth}{!}{                                    % used for centering table
\begin{tabular}{ccccccccccc}          % centered columns (4 columns)
\hline\hline                        % inserts double horizontal lines
  Run label  & Resolution $N_R\times N_\varphi$&${\it St}$ & $\epsilon$ & $Q_g$ &  $\alpha$ & $D$ & $c_p^2/c_g^2$ & $kH_g$& $\sigma$  & Outcome \\ % table heading
\hline                                   % inserts single horizontal line

L1 &$2048\times 2048$& $0.1$ & $0.1$ &  $3$ &  & $0$ & $10^{-2}$ & 4 & $3.2\times 10^{-3}$ & dust clumps\\
L2 &$1024\times 1024$ &$0.01$ & $0.1$ &  $3$ &  & $10^{-4}$ & $0$   &  4 & $2\times 10^{-4}$&   m=1 spiral \\
R1  & $2048\times 2048$&$1$ & $0.01$ &  $3$ & $10^{-4}$  & Eq.\ref{eq:d} & Eq.\ref{eq:cp}& 10 & $1.4\times 10^{-2}$&  dust clumps\\
%L3 & $10$ & $0.01$ &  $5$ &  & $0$ & $10^{-4}$ & 17 & $3\times 10^{-3}$& $\sim 10^3$ & RWI\\
%R2  & $1$ & $0.01$ & $5$ & $10^{-4}$ & Eq.\ref{eq:d} &  Eq.\ref{eq:cp}& 6 & $1.6\times 10^{-3}$& $\sim 1$ & RWI\\
R2  & $2048\times 2048$&$0.1$ & $0.1$ & $7$ & $10^{-4}$ & Eq.\ref{eq:d} &  Eq.\ref{eq:cp}& 10 & $7\times 10^{-3}$  & dust clumps\\
 %R4  & $0.01$ & $1$ &  $7$ &$10^{-4}$ & Eq.\ref{eq:d}&  Eq.\ref{eq:cp}&  12 &$1.6\times 10^{-2}$ & $\sim 10$ & RWI\\
 %R5  & $0.1$ & $0.01$ & $7$ & $10^{-5}$ & Eq.\ref{eq:d}& Eq.\ref{eq:cp} &10& $6\times 10^{-4}$& \\
 R3  & $2048\times 2048$&$0.01$ & $0.1$ & $7$ & $10^{-5}$ & Eq.\ref{eq:d}&  Eq.\ref{eq:cp}& 14 &$1.1\times 10^{-3}$ & dust clumps \\
 R4  &$2048\times 2048$ &$0.1$ & $0.01$ & $5$ &$10^{-5}$ & Eq.\ref{eq:d}&  Eq.\ref{eq:cp}&15& $2.1\times 10^{-3}$ &dust clumps \\
 R5  & $2048\times 2048$&$0.01$ & $0.1$ &  $5$ & $10^{-5}$ & Eq.\ref{eq:d}&  Eq.\ref{eq:cp}&20 & $3.1\times 10^{-3}$& dust clumps  \\
 R6  &$2048\times 2048$ &$0.01$ & $0.01$ &  $3$ & $10^{-5}$ & Eq.\ref{eq:d}& Eq.\ref{eq:cp} &4 & $2.5\times 10^{-5}$ & m=1 spiral \\
\hline                                             %inserts single line
\end{tabular}
\end{table*}

In this section, we first test the capability of our code to recover the expected  linear growth rates of the secular gravitational instability. This is achieved by employing two different test problems. 

 The first model, hereafter labelled as $L1$, has $Q_g=3$, $\epsilon=0.1$, ${\it St=0.1}$, $D=0$, $c_p^2/c_s^2=10^{-2}$ and has been used by Tominaga (2018) to study the non-linear evolution of the axisymmetric SGI. For this model, the wavenumber of the most unstable mode is such that $k\sim 3-4H_g^{-1}$ with $H_g$  the gas disc scaleheight and the maximum growth rate is $\sigma\sim 3\times 10^{-3}\Omega$.  The second model, labelled in the following as $L2$,   has  $Q_g=3$, $\epsilon=0.1$, ${\it St=0.01}$, $D=10^{-4} c_s^2 \Omega^{-1}$, $c_p^2/c_s^2=0$. It corresponds to the model considered in Takahashi \& Inutsuka (2014) and for which the most unstable mode is again  $kH_g\sim 3-4$ and the growth rate is $\sigma\sim 2\times 10^{-4}\Omega$. Parameters of  these two test runs can also be found in Table \ref{table1}, where we summarize the parameters that were used in each of the simulations we performed.

For each of the two models, the simulations are initialized with axisymmetric perturbations that correspond to the eigenvectors of the most unstable mode. These initial perturbations have amplitude $A=10^{-4}f_s$ and are superimposed to the initial surface density and velocity components given in Sect. \ref{sec:init}.  Here, $f_s$ is defined as: 
\begin{equation}
f_s=\exp\left[\frac{-(R-R_0)^2}{2\sigma^2}\right]
\end{equation}

where $\sigma=4/k$, and is employed to minimize the influence of boundaries.
 For these two tests of the SGI linear growth, the adopted numerical resolution  is such that the most unstable wavelength is resolved by  $\sim 128$ grid cells.

Fig. \ref{fig:linear} shows, as a function of time and for both models,  the maximum deviations from initial state of the dust surface density, gas surface density and dust-to-gas ratio. Here, the maximum deviation of each fluid variable has been measured for distances in the range $0.8<R<1.2$. Exponential growth of the perturbations is observed to arise at early times, with growth rates that are in good agreement with those expected from linear theory, and such that  the dust-to-gas ratio remains almost constant.  For model L2, the oscillations that can be observed at Times $\lesssim 2000$ arise because of overstability, which is also expected from linear theory (Takahashi \& Inutsuka 2014). This shows than the code can satisfactorily capture the SGI and reproduce analytical growth rates. At later times, however, a significant increase of the growth rate is observed, although this is found to arise due to different processes in each  model. \\

\subsection{Non-linear evolution}

\subsubsection{Ring fragmentation into dusty clumps}

For model L1, the steep increase of the perturbation amplitudes is found to be related to  the gravitational collapse of the dust rings, consistently with the results of Tominaga (2018). This is demonstrated in Fig. \ref{fig:toomre_b} where we show  the time evolution of the Toomre parameter of the gas and of the dust.  The Toomre parameter of the dust decreases below the gravitational stability limit at Time$\sim 391$, which coincides with the time of fast increase of the density perturbation amplitudes. Gravitational instability of the dust component is also suggested by looking at maps of the relative dust density perturbation in Fig. \ref{fig:2d_modeleb}.  Comparing the Time=395 and Time=397 panels,   we clearly see that the dust rings become much thinner as the gravitational collapse proceeds. Interestingly, we find that these dusty rings ultimately break into clumps. It appears that the formation of these dust clumps results from the RWI, which is expected to develop at a 
maximum of the potential vorticity (PV), which is defined as (Lin \& Youdin 2017): 

\begin{equation}
{\cal \nu}=\frac{\kappa^2}{2\Omega(\Sigma_g+\Sigma_p)}\left(1+\frac{\Sigma_p}{\Sigma_g}\right)^2
\end{equation}

 \begin{figure}
\centering
\includegraphics[width=\columnwidth]{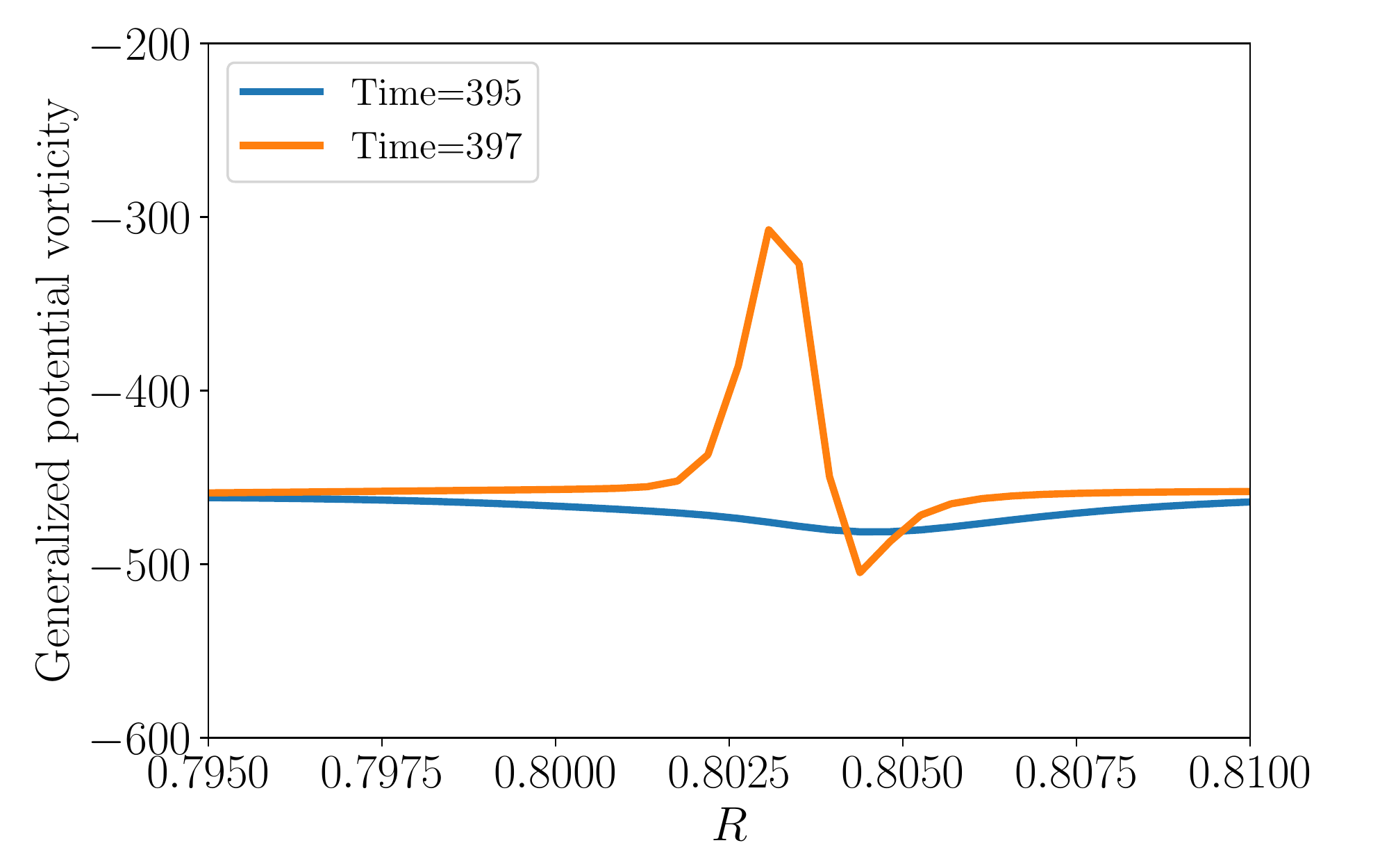}
\caption{Azimutally averaged profile of the generalized potential vorticity for model L1 at different times.}
\label{fig:gpv_l1}
\end{figure}

where $\kappa$ is the epicyclic frequency. Fig. \ref{fig:gpv_l1} shows the generalized PV profile at different times for model L1. We see that there is a clear trend for the 
bump in the PV profile to increase as the collapse of the dust ring proceeds, which appears therefore as a preferred location to trigger the RWI. By examining at Time=404 contours of the Rossby number $Ro$, which is defined as :

\begin{equation}
Ro=\frac{{\bf e_z}\cdot(\nabla\wedge({\bf v}-R\Omega{\bf e_\varphi}))}{2\Omega}
\end{equation}

\begin{figure}
\centering
\includegraphics[width=\columnwidth]{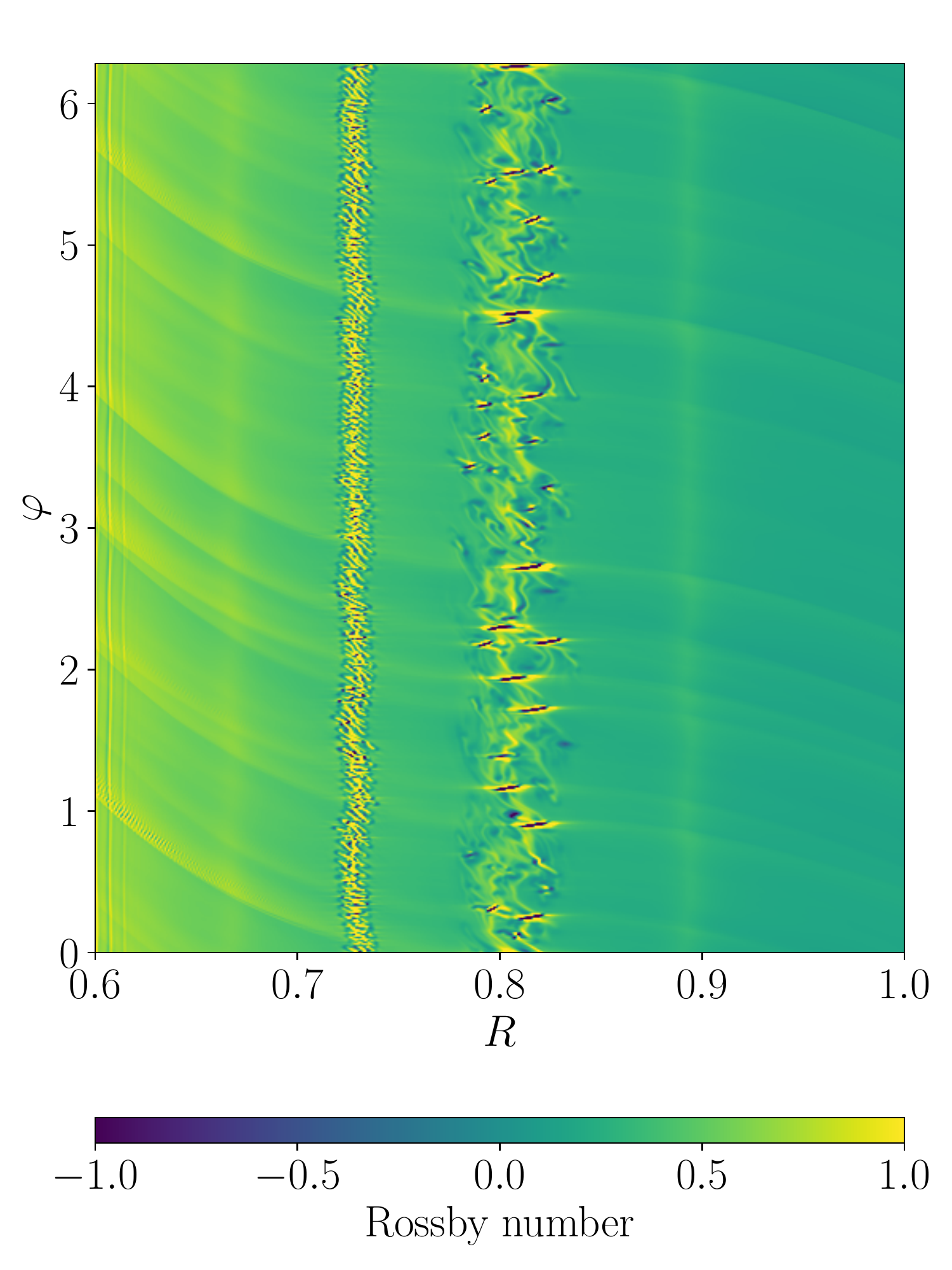}
\caption{For Model L1, contours of the Rossby number at Time=404. }
\label{fig:gpv2d}
\end{figure}

 and which are presented in Fig. \ref{fig:gpv2d}, we can confirm that the  dust clumps that are formed indeed correspond to anticyclonic vortices resuting from the RWI. It is interesting to note  that the larger ring width that can be observed at the Time=404 panel of  Fig. \ref{fig:2d_modeleb} suggests significant gravitational interactions between these clumps, which may limit further growth of these clumps. \\

\begin{figure*}
\centering
\includegraphics[width=\textwidth]{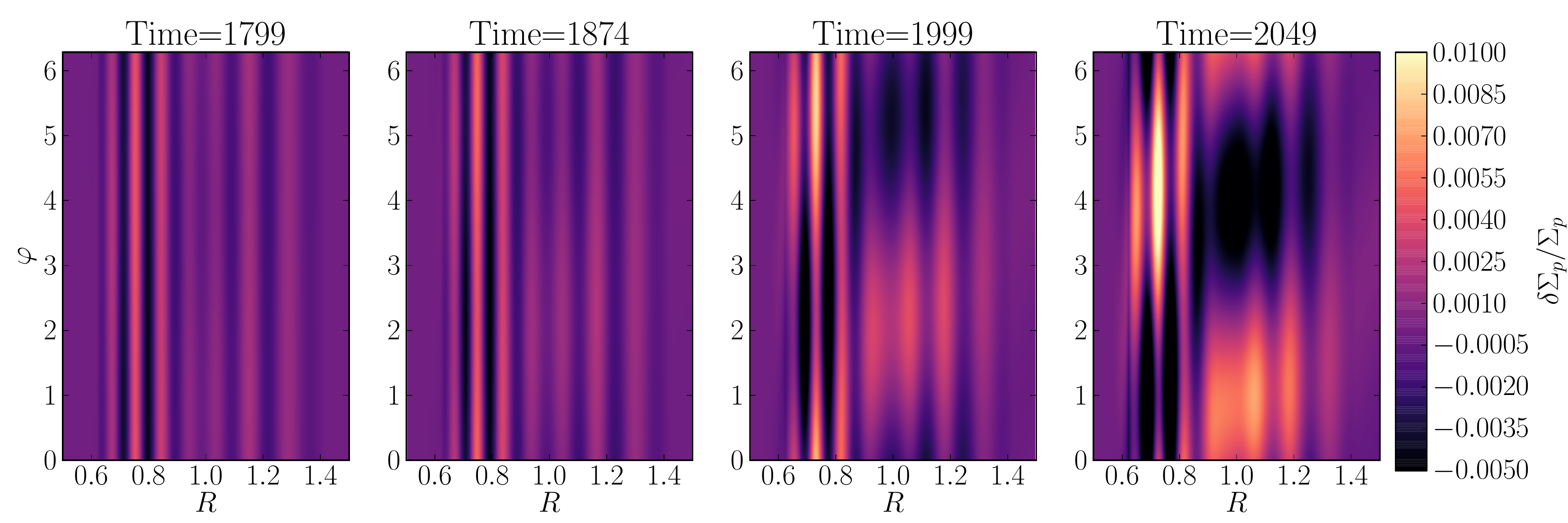}
\includegraphics[width=\textwidth]{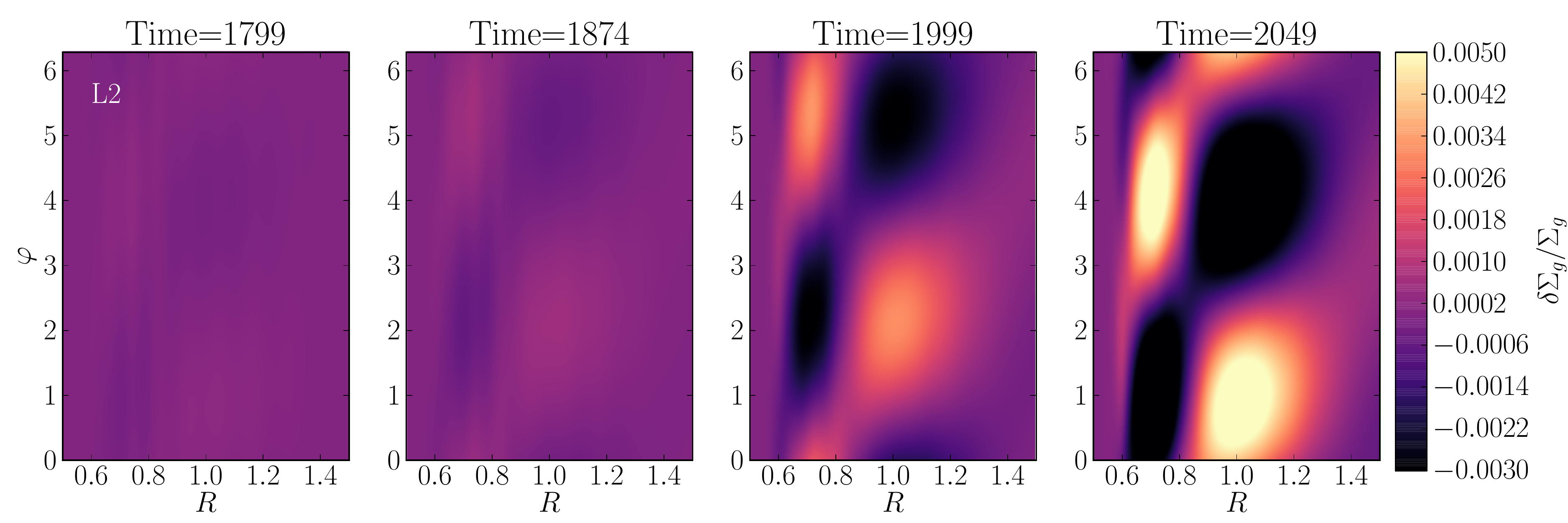}
\caption{{\it Top:} Relative dust density perturbation at different times for model L2.  {\it Bottom:} same but for the gas component.}
\label{fig:fig_modelea}
\end{figure*}

\subsubsection{Non-axisymmetric instability driven by dust feedback}

A period of fast growth of the perturbations amplitudes followed by non-linear saturation is also observed for model $L2$, although a different mode of evolution is obtained in that case. As illustrated in Fig. \ref{fig:fig_modelea}, a $m=1$ spiral wave is a found to  develop in the gas disc   with a growth rate  $\gamma\sim 0.003 \Omega$.  To explore possible origins of this instability, we have conducted a number of test simulations whose results may be summarised as follows:
\begin{itemize}
\item Simulations in which the indirect term of the gravitational potential was discarded also  resulted in the development of the one-armed spiral. In  previous numerical simulations of vortices in self-gravitating discs, it has been found that for Toomre parameters $Q_g\lesssim 3$, a $m=1$ non-axisymmetric mode can develop as the barycentre of the system is significantly shifted away from the central star (Pierens \& Lin 2018). Hence, the origin of the instability is  not related to  the movement of the central star  nor to the SLING mechanism (Adams et al. 1989).
\item It has been suggested that one armed spirals might be excited in the presence of a forced temperature gradient  (Lin 2015). This arises because a temperature gradient leads to angular momentum exchange between the wave and the background disc, such that a torque is exerted on the $m=1$ perturbation.  This $m=1$ spiral 
wave is expected to be confined between Q-barriers which occur at radii $R_{Qb}$ that are defined by (Lin 2015):
\begin{equation}
Q^2(R_{Qb})\left[1-\nu^2(R_{Qb})\right]=1
\end{equation}

where 
\begin{equation}
\nu=\frac{\omega-\Omega}{\kappa}
\end{equation}
It is also expected to grow with a corresponding growth rate given by (Lin 2015):
\begin{equation}
\gamma \sim \frac{1}{2}\frac{dc_s^2}{dR}\frac{k}{\Omega}
\end{equation}
 with $kR\sim \frac{1}{hQ_g} $.
Here we find  that the $m=1$ mode has pattern speed $\Omega_p=\omega\sim 1.017 \Omega$, resulting in Q-barriers that would be located at $R\sim 0.23$ and $R\sim 1.5$.  Looking at Fig. \ref{fig:fig_modelea}, a by-eye determination of the region where the $m=1$ spiral wave can propagate  is in  reasonable agreement with the previous estimation,  as the instability tends to operate in the entire numerical domain.  We note in passing that the fact that Q-barriers are located outside of the numerical domain exclude the SWING and WASER amplifications as potential mechanisms for the growth of the $m=1$ mode. Both processes result from the leakage of density waves across Q-barriers and  have been invoked by Noh et al. (1991) to explain the $m=1$ instability they observed in their simulations of non-axisymmetric instabilities of dusty-gas discs.    Moreover, we find that the mechanism presented in Lin (2015) would  give an estimating growth rate of  $\sim 0.01 \Omega$,  namely larger to that found numerically. Furthermore,  test simulation with a zero temperature gradient showed growth of the $m=1$ mode,  which discards this mechanism as a possible explanation of the instability.
\item Growth of the $m=1$ spiral wave failed in simulations where the effect of the dust feedback was discarded, which emphatises the important role of the angular momentum exchange between the gas and particles in triggering the instability. To clearly  unveil the important role of dust backreaction we computed the torques $T_{BG}$ and $T_{FB}$ 	associated respectively with the background state of the gas disc and the dust feedback. These are given respectively by (Lin 2015):
\begin{equation}
T_{BG}=\frac{1}{2}\int_{R_{in}}^{R_{out}}\frac{dc_s^2}{dR}k\Sigma_g|\xi_R|^2 2\pi R dR
\label{eq:tbg}
\end{equation}

where $|\xi_R|$  is the radial component of the Lagrangian displacement (Papaloizou \& Pringle) and which can be approximated by $|\xi_R|\sim |v_R|/\Omega$ as $\sigma << \Omega$, and:

\begin{equation}
T_{FB}=-\frac{1}{2}\int_{R_{in}}^{R_{out}}\frac{\Sigma_p}{\tau_s}(v_\varphi-V_\varphi) 2\pi R^2 dR
\label{eq:tfb}
\end{equation}

\end{itemize}

\begin{figure}
\centering
\includegraphics[width=\columnwidth]{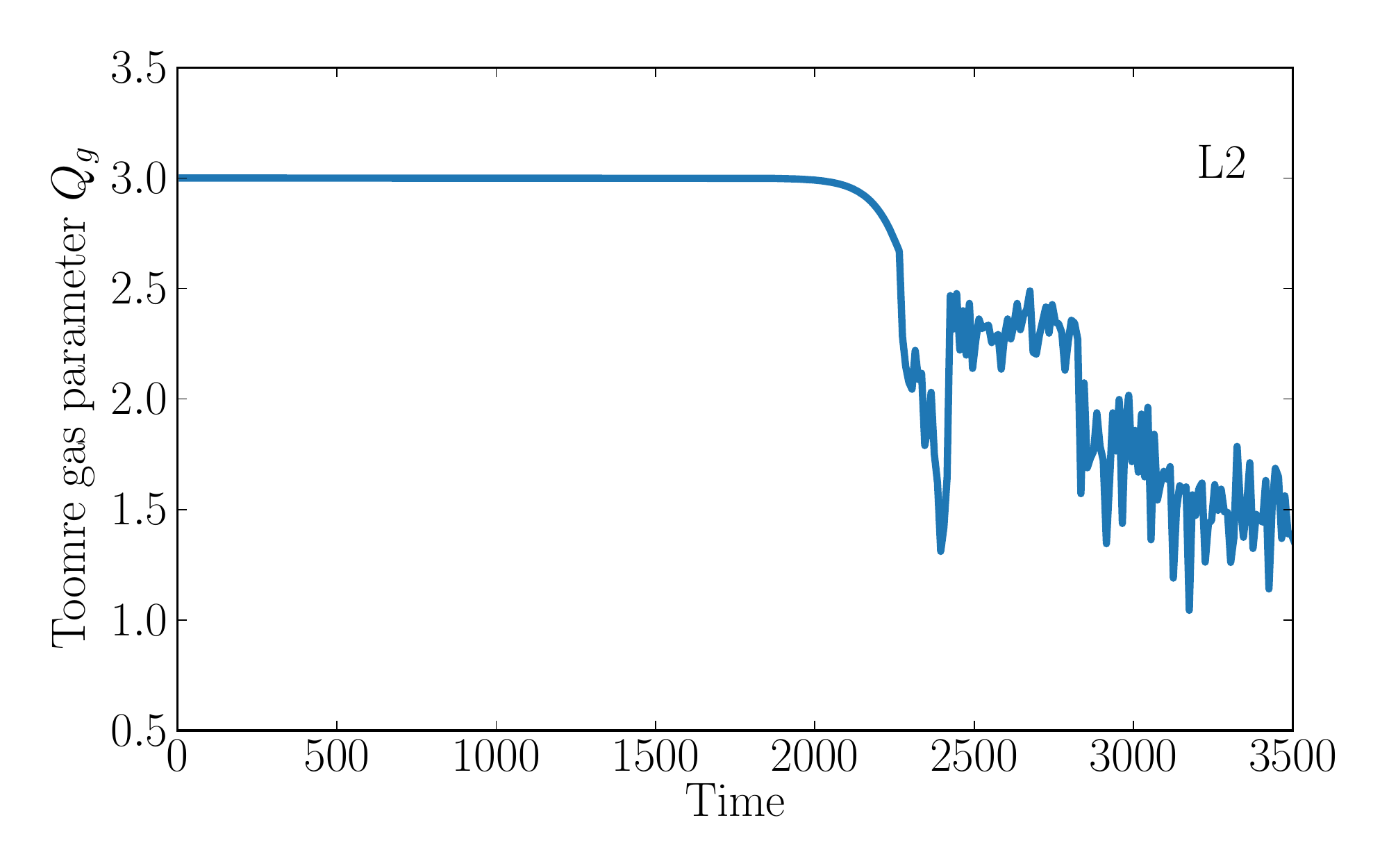}
\caption{For Model L2, temporal evolution of the gas Toomre stability parameter.  }
\label{fig:toomre_a}
\end{figure}

  Examining the evolution of both quantities as a function of time reveals that in comparison with the background torque,  the torque arising from dust feedback is significantly larger. This leads us to interpret the formation of the one-armed spiral as simply a non-axisymmetric instability in the gas disc triggered by the effect of dust feedback.  Although in a purely gaseous disc non-axisymmetric instability occurs for $Q_g\sim 2$ typically,   it can be reasonably expected that due to dissipation resulting from gas-dust interaction, non-axisymmetric instabilities can arise for even larger values of $Q_g$. This is also suggested by  Fig. \ref{fig:toomre_a} which shows the temporal evolution of $Q_g$ . We see that $Q_g\sim 3$ at the point in time when the instability develops, namely at Time$\sim 2000$ according to Fig.  \ref{fig:fig_modelea}, which thereby corroborates our previous interpretation.

\begin{figure}
\centering
\includegraphics[width=\columnwidth]{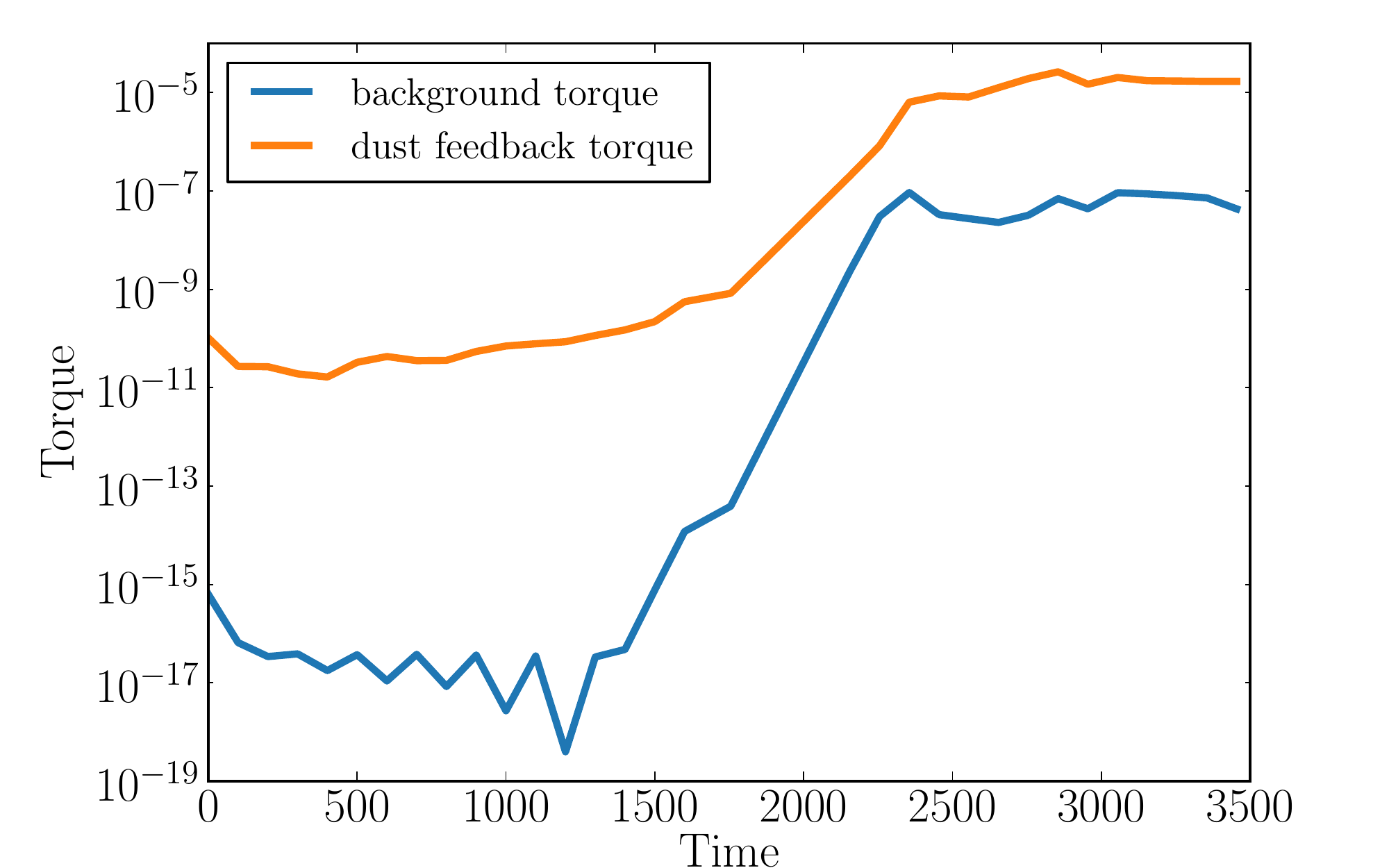}
\caption{For Model L2, temporal evolution of the background torque and torque due to dust feedback and given by Eqs.  \ref{eq:tbg} and \ref{eq:tfb} respectively. }
\label{fig:torque}
\end{figure}

\subsection{Dust concentration}
\label{sec:concentration}

In this section, we discuss the results of simulations that examine the non-linear evolution of the SGI in models where both the dust diffusion coefficient and velocity dispersion are 
determined self-consistently from Eqs. \ref{eq:d} and \ref{eq:cp}. These simulations are labelled as $R1,...,R6$ in Table \ref{table1}. In all but model $R6$, dust ring amplification was found with  non-axisymmetric breakdown due to the RWI occuring subsequently,  similarly to  what has been found for model L1. The  evolution in model $R6$, however, which has $Q_g=3$ and a linear growth timescale of $\sim6\times 10^4$ orbits, is consistent with the model $L2$ presented above, showing growth of a $m=1$ spiral wave driven by the gas component.

\begin{figure*}
\centering
\includegraphics[width=0.49\textwidth]{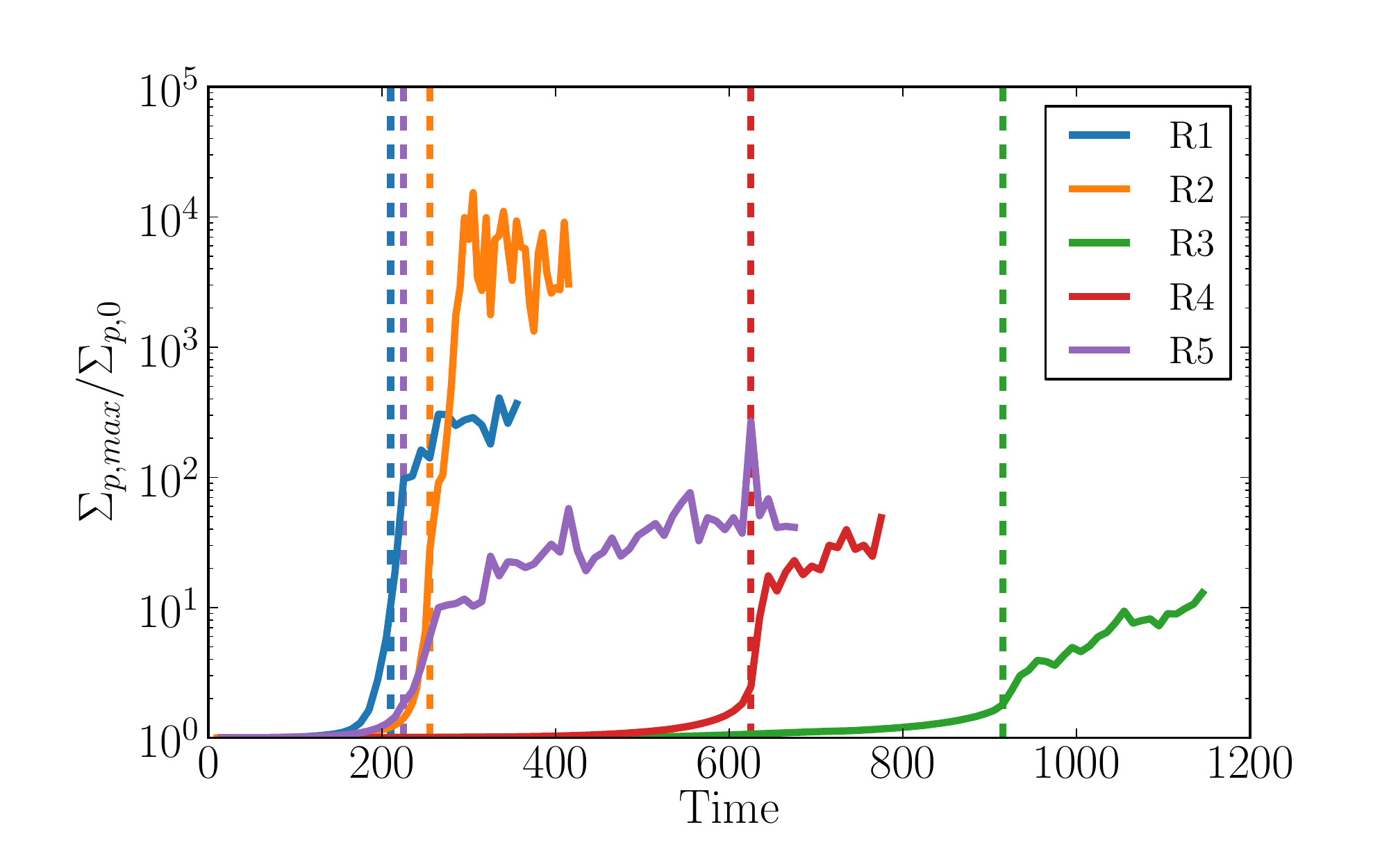}
\includegraphics[width=0.49\textwidth]{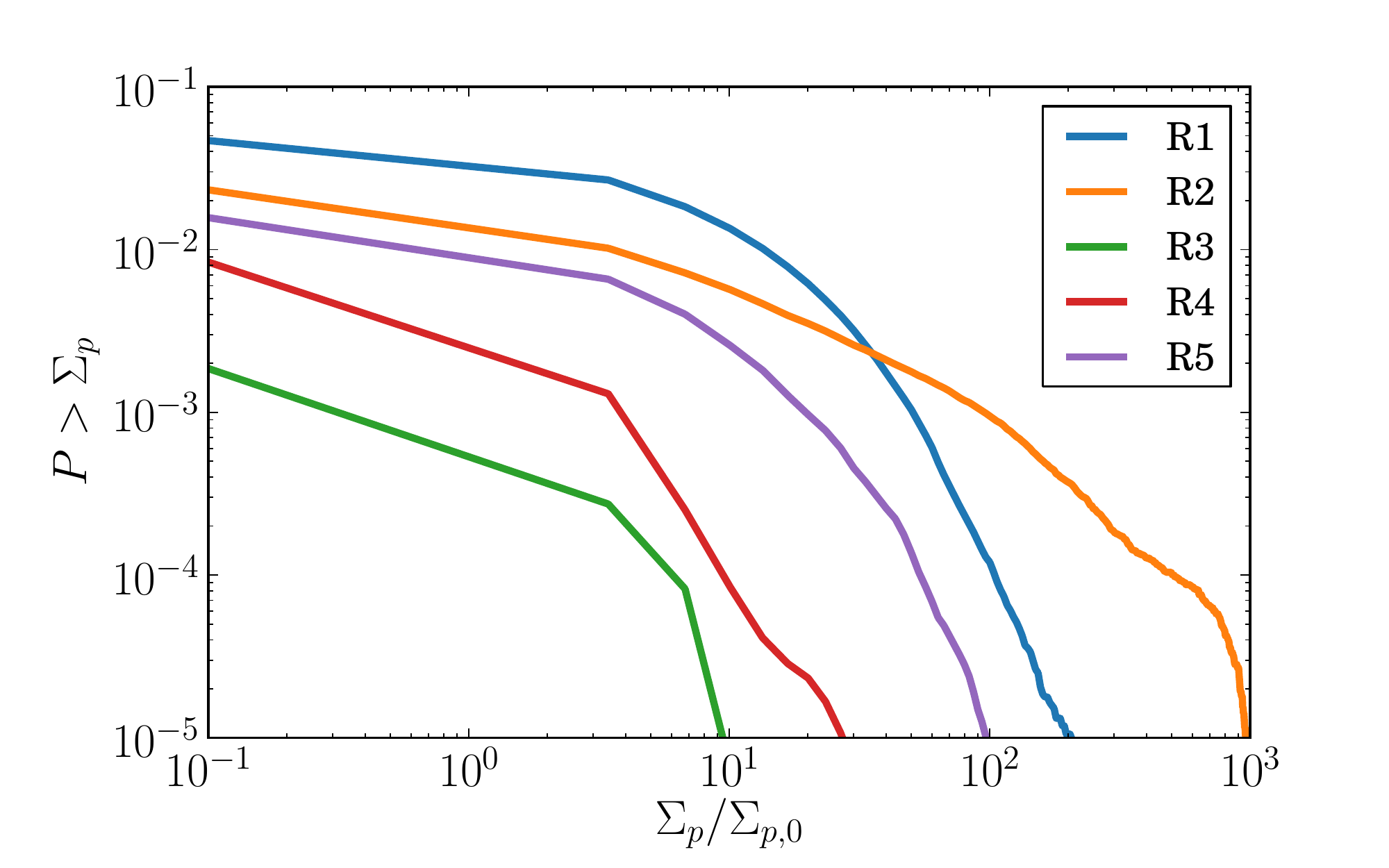}
\caption{{\it: Left:}Time evolution of the maximum dust density for models $R1,...,R5$. For each model, the vertical dashed line marks the time when vortices are formed through the RWI. {\it Right:} Cumulative particle density distribution. }
\label{fig:dustmax}
\end{figure*}

The left panel  of Fig. \ref{fig:dustmax} shows the time evolution of the maximum dust density for models $R1,...,R5$. A general picture that emerges from these simulations is a clear trend for the maximum dust density to scale with the linear growth rate of the SGI. For instance, the relative maximum dust density $\Sigma_{p,max}/\Sigma_{p,0}$, where $\Sigma_{p,0}$ is the inital particle surface density,  is $\sim 10^4$ for model $R2$ which has $\sigma\sim 7\times 10^{ -3} \Omega$, whereas $\Sigma_{p,max}/\Sigma_{p,0}\sim 30$ for model $R4$ for which $\sigma\sim 2\times 10^{ -3} \Omega$.   This is also illustrated by the right panel of Fig. \ref{fig:dustmax} which displays the cumulative particule distribution, which is  computed by using 300 dust density bins
 in the range $[0.1,1000]\Sigma_{p,0}$ and then counting the cells with dust density above a certain value. Model $R1$ is the one anomaly, since it  resulted in a weaker clumping compared than model $R2$, although the corresponding linear growth rate is higher. Looking back at the CPD, this model however exhibits a larger number of cells with $ 1 \lesssim \Sigma_{p,max}/\Sigma_0 \lesssim 50$ compared to model $R2$. \\
 
 \begin{figure}
\centering
\includegraphics[width=\columnwidth]{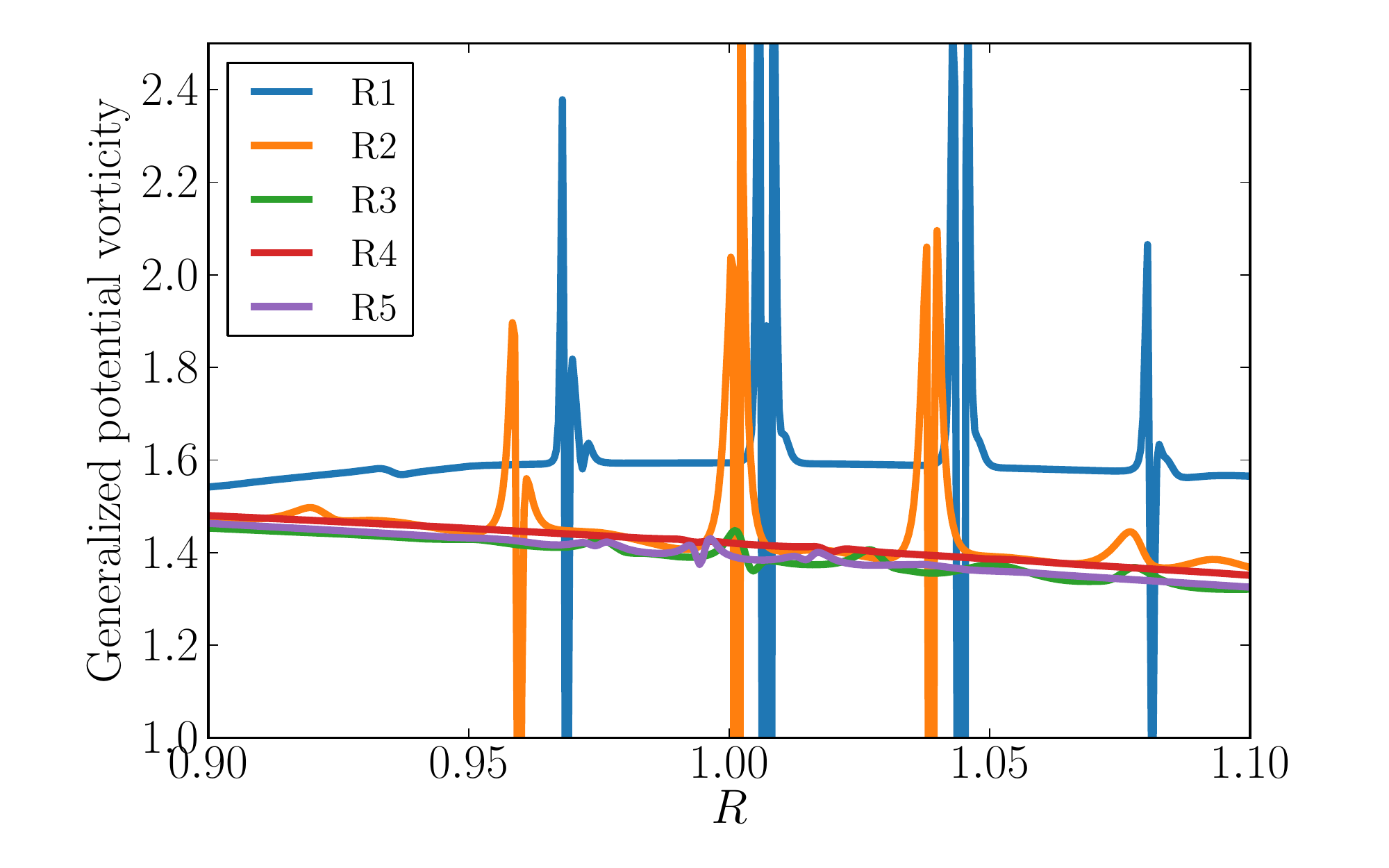}
\caption{Azimutally averaged profile of the generalized potential vorticity for models R1,...,R5.}
\label{fig:gpv}
\end{figure}

 The dependence of particle clumping efficiency with the value of the SGI growth rate is likely to arise because higher growth rates tend to result in rings with smaller radial width and   stronger bump in the generalized PV profile, as can be seen in Fig. \ref{fig:gpv} which shows the PV profile at a point in time before non-axisymmetric breakdown into vortices occurs. This phenomenon is in rather good  agreement with the results of  Pierens et al. (2019) who found that for vortices forming at a planet separatrix, the evolution is somewhat more violent for particles for which the bump in the PV profile located at the planet separatrix is the strongest. Interestingly, an analogy can also  be drawn between our results and the ones of Hammer et al. (2017) in purely gas discs. These authors indeed found that in the case of vortices forming at the gap edge of a giant planet, slowly growing-planets tend to produce weaker and more elongated vortices, in contrast to shorter growth timescales which lead to more concentrated vortices. 
 
\begin{figure*}
\includegraphics[width=0.33\textwidth]{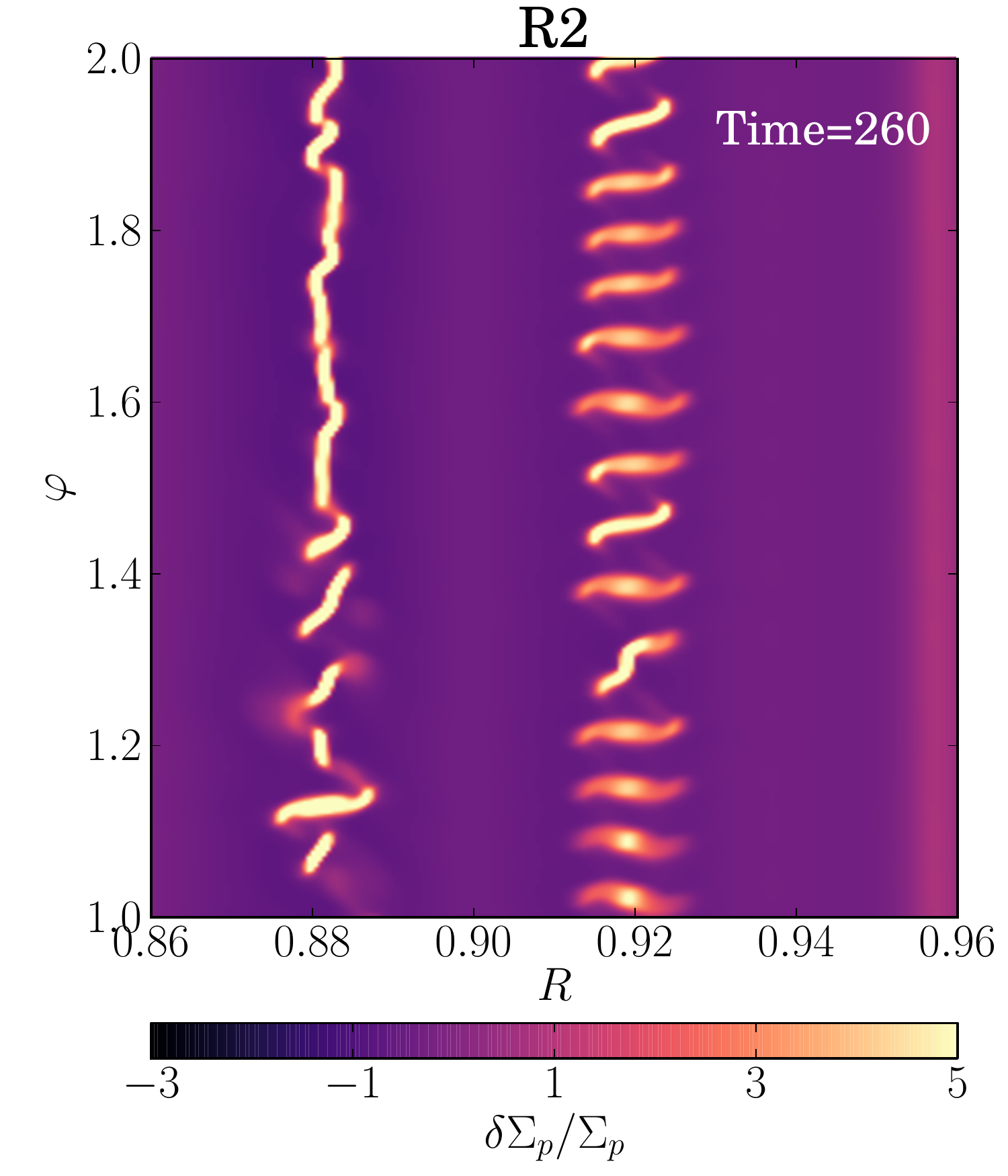}
\includegraphics[width=0.33\textwidth]{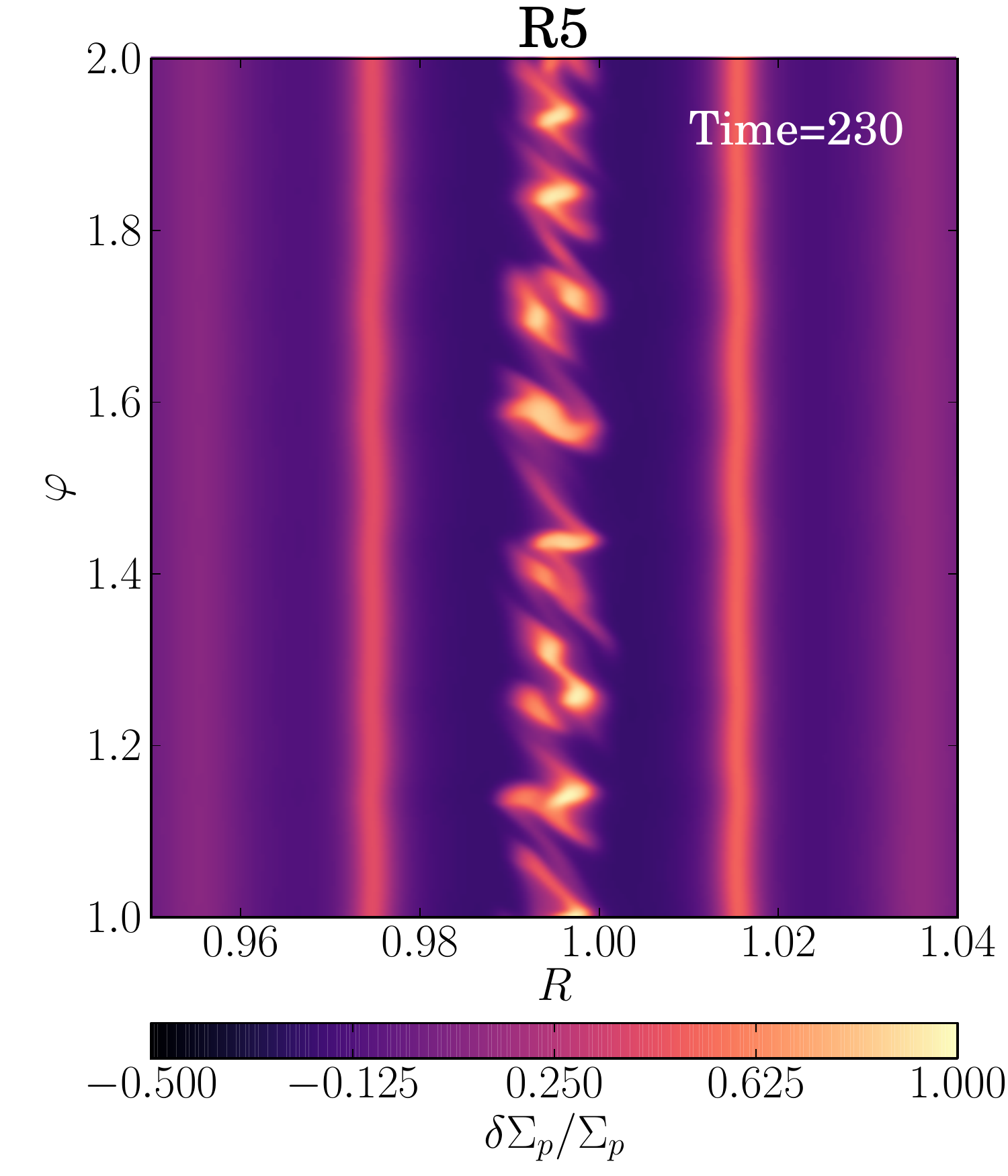}
\includegraphics[width=0.33\textwidth]{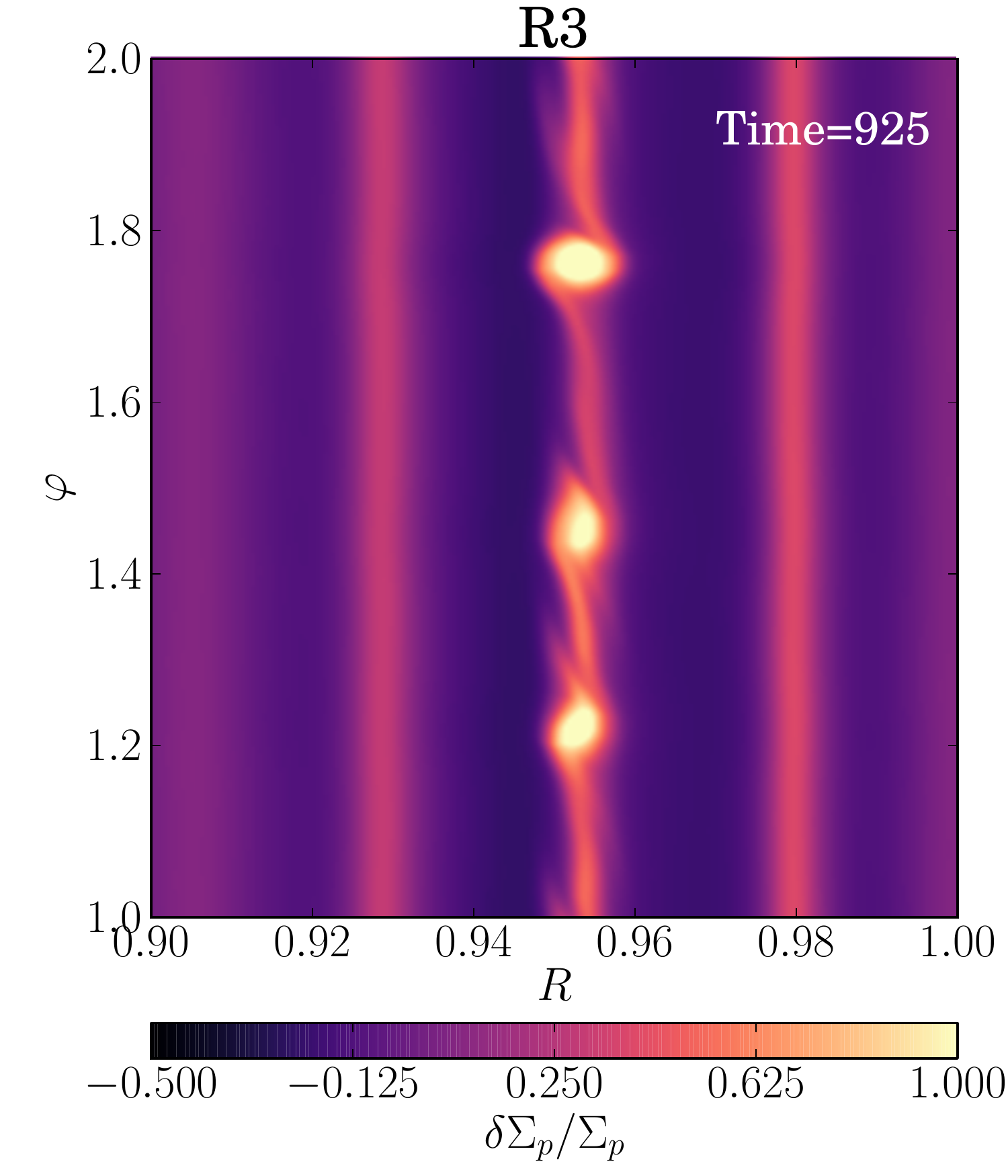}
\caption{Relative dust density perturbation for models R2, R3, R5 at a point in time when dusty vortices are formed.}
\label{fig:vortices}
\end{figure*}

Fig. \ref{fig:vortices} shows the relative dust density for models R2, R3, R5 at times corresponding to the onset of the RWI. More vortices are produced in models with higher linear growth rates and that present a sharper gradient in the PV profile and therefore a more vigorous instability.  One can also notice a clear tendency for the azimutal extent of these vortices to increase as the growth rate  decreases. Again, this is consistent with the results of Hammer et al. (2017) who found that slowly-growing planets form vortices that are up to twice as elongated compared to planets to planets with shorter growth timescale. \\

The  process described above can in principle lead to the formation of gravitationally bound objects if the local particle density $\rho_p$ exceeds the Roche density which is given by:

\begin{equation}
\rho_{Roche}=\frac{9\Omega^2}{4\pi G}
\end{equation}

Using the fact that $H_p\sim \left(\frac{\alpha}{\it St}\right)^{1/2}H_g$ for $\alpha << {\it St}<<1$ and $\rho_p\sim \frac{\Sigma_p}{\sqrt{2\pi} H_p}$, we find $\rho_p>\rho_{Roche}$ at saturation for models R1,R2,R5 whereas  $\rho_p\lesssim\rho_{Roche}$ for models R3, R4. As suggested by the slight increase in $\Sigma_{p,max}/\Sigma_{p,0}$ that is observed at later times for these two models in Fig. \ref{fig:dustmax}, the dusty vortices that emerge from the RWI further accumulate dust particles such that it can not be excluded that the Roche density will be eventually reached. We note however that the maximum dust-to-gas ratio inside these vortices, which is $\epsilon_{max}\sim 10$ for model R3 and $\epsilon_{max}\sim 1$ for model $4$ is above the threshold for which the vortex might be destroyed  by the effect of dust feedback (Fu et al. 2014; Crnkovic-Rubsamen et al.; Raettig et al. 2015). Higher resolution simulations would be required to examine potentially important effects that might occur within the vortex core.

\section{Conclusions}

In this paper, we examined the non-linear evolution of the Secular Gravitational Instability (SGI) using two-fluid hydrodynamical simulations with self-gravity. Previous work focusing on the non-linear phase of the SGI was restricted to axisymmetric simulations such that the important issue of the formation of planetesimal clumps through the SGI could not be addressed. We considered disc models with zero pressure gradient in which were added initial perturbations corresponding to the eigenfunction of most unstable mode obtained from an axisymmetric linear analysis.  After a linear growth phase  during which the amplitude of the ring perturbation initially increases with a growth rate that is found to be in good agreement with the linear analysis, we obtain two possible modes of evolution depending mainly on the value for the initial linear growth rate: 
\begin{itemize}
\item In discs with Toomre gas parameter $Q_g\lesssim 3$ and where the growth rate of SGI  is smaller than $\sigma \gtrsim 10^{-5}-10^{-4}\Omega$, dissipation resulting from dust feedback introduces a $m=1$ spiral wave in the gas disc which can subsequently trap solids until a dust-to-gas ratio $\epsilon \sim 1$ is achieved. This result implies that due to dust-gas interaction, the critical Toomre parameter for which spiral waves can emerge from the gas disc tends to be higher than in a purely gaseous disc.  \\ 
\item For smaller linear growth rates, the dust ring  is found to undergo gravitational collapse until the  bump in the solid surface density profile becomes strong enough to trigger the formation of dusty clumps. These clumps are actually dusty vortices that emerge because a bump in the surface density implies an extremum in the  generalized potential vorticity (PV), such that it appears as a favored location for the growth of the Rossby Wave Instability (RWI).  Because higher growth rates lead to sharper PV gradients, there is a clear tendency for the RWI to be more vigorous in that case, with enhancements in the dust concentration that can be as high as $\sim 10^3-10^4$ and which correspond to bound particle clumps.
\end{itemize}
An important caveat of this work is certainly the zero pressure gradient disc model that we adopted, possibly resulting in ovestimates of the maximum dust density that was achieved in the simulations. Although it has been demonstrated that taking into account the effect of dust radial drift does not significantly impact the SGI growth rate, an additional constraint for the SGI to produce dense dust rings  is that the growth timescale should be shorter than the drift timescale  (Tominaga et al. 2020). Moreover, since the secular GI does not create significant gas substructures in that case (Tominaga et al. 2020), it is not clear whether or not the first mode of evolution that we found and corresponding to the formation of a $m=1$ spiral wave remains possible in discs with non zero pressure gradients. We will examine the issue of ring fragmentation in discs with various surface density profiles in a future study.

\section*{Acknowledgments}
Computer time for this study was provided by the computing facilities MCIA (M\'esocentre de Calcul Intensif Aquitain) of the Universite de Bordeaux and by HPC resources of Cines under the allocation A0090406957 made by GENCI (Grand Equipement National de Calcul Intensif).  We thank Min-Kai Lin for fruitful discussions.

\section*{Data Availability}

The data underlying this article will be shared on reasonable request to the corresponding author.

\appendix

\section{Diffusion test}
\label{sec:diffusion}

\begin{figure}
\centering
\includegraphics[width=\columnwidth]{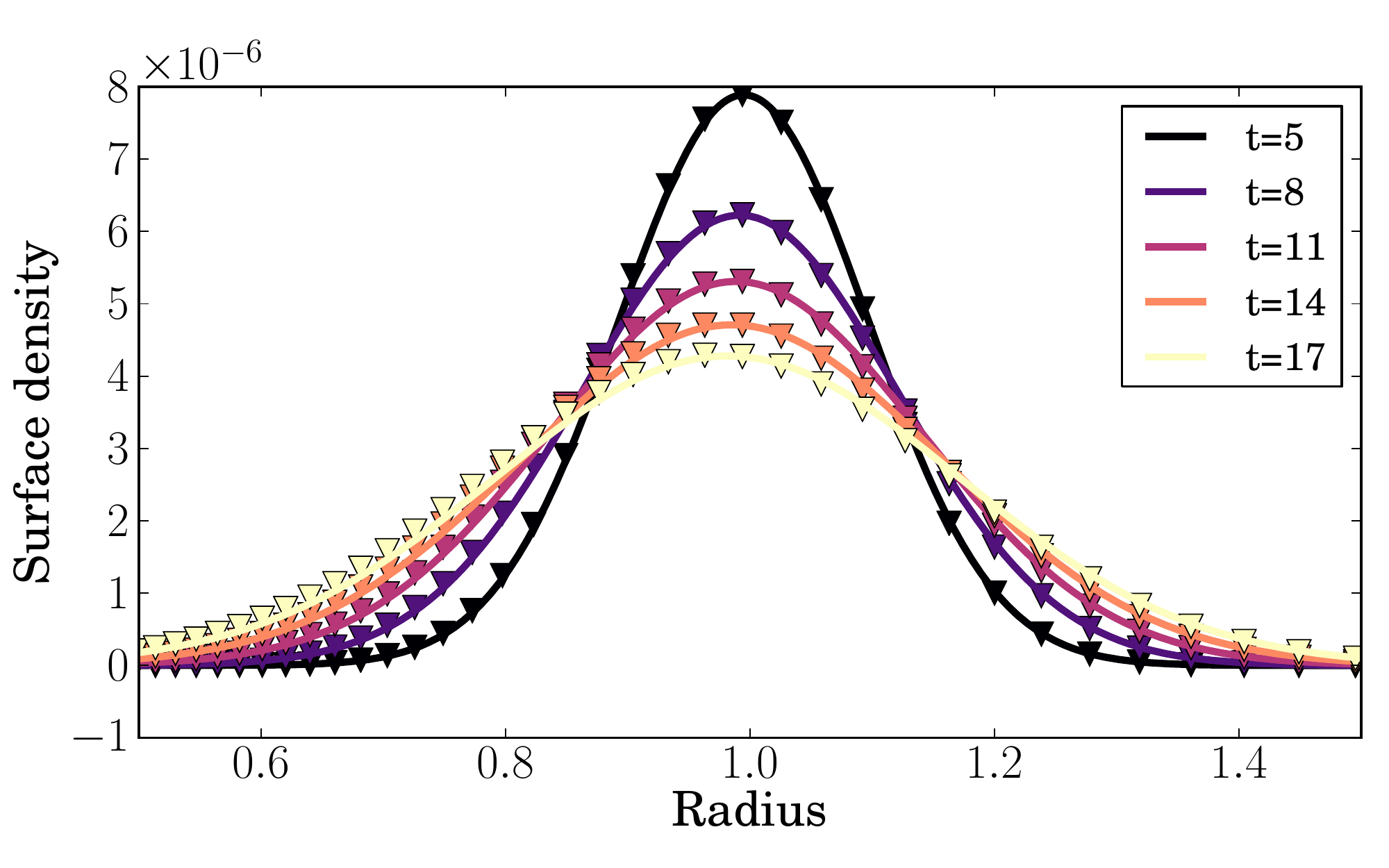}
\caption{Temporal evolution of the dust density profile for the test of the dust diffusion implementation described in Appendix  \ref{sec:diffusion}}.
\label{fig:test_settling}
\end{figure}

Here, we show that our implementation of diffusion  is correct and accurately reproduces the evolution of an axisymmetric dust ring that vicously spreads in a gas disc. To this aim, we repeat the test  of dust diffusion in the radial direction presented in Weber et al. (2019), with an initial density distribution $\rho(r,t=2)$, where $\rho(r,t)$ is given by: 
\begin{equation}
\rho(r,t)=\frac{mr_0}{2Dt}\exp\left(-\frac{r^2+r_0^2}{4Dt}\right)I_0\left(\frac{rr_0}{2Dt}\right)
\end{equation}
where $D$ is the dust diffusion coefficient,  $m$ is a normalization factor, $r_0$ corresponds to the location of the density peak,  and $I_0$ the modified Bessel function of the first kind. Similarly to Weber et al. (2019), we set $D=10^{-3}$, $m=2\times 10^{-6}$, $r_0=1$, $r\in[0.1,2.5]$ and use $N_r=1024$ radial grid cells. The results of the test are displayed in Fig. \ref{fig:test_settling}, where we see that the expected analytical estimate is recovered.

\section{Diffusion modelling using diffusion pressure}

\begin{figure*}
\centering
\includegraphics[width=\textwidth]{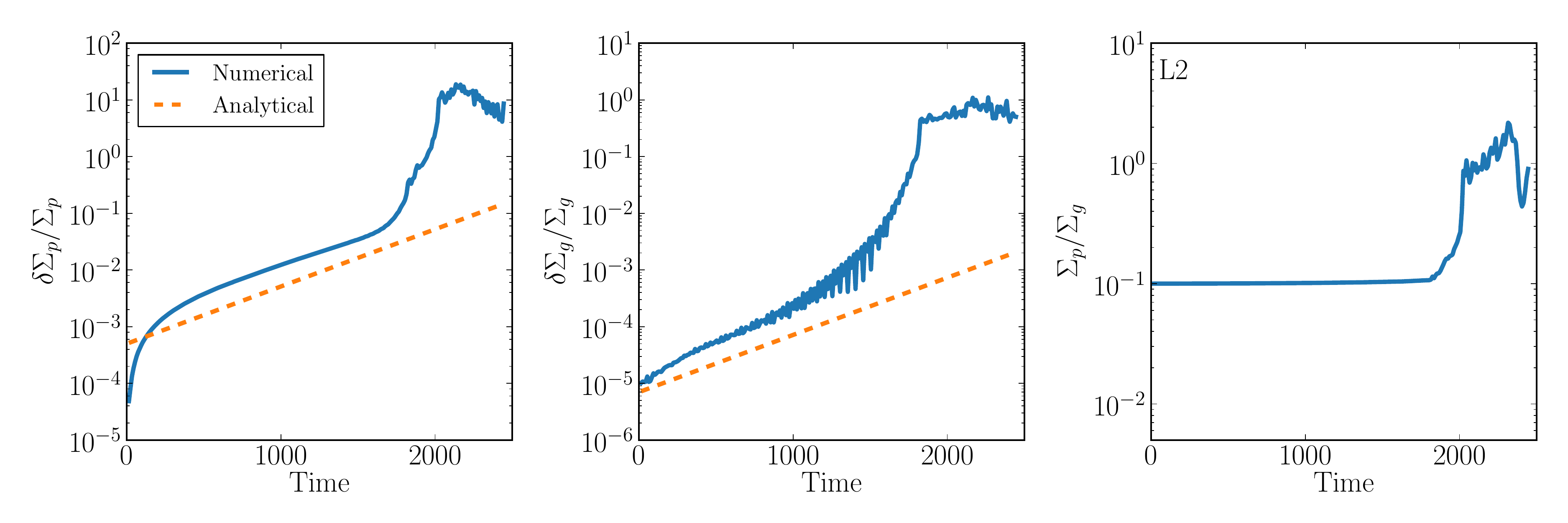}
\caption{From left to right, temporal evolution of the dust surface density, gas surface density, and dust-to-gas ratio for model L2 but in which  a diffusion pressure is employed for diffusion modelling.}
\label{fig:linear_2}
\end{figure*}

\begin{figure*}
\centering
\includegraphics[width=\textwidth]{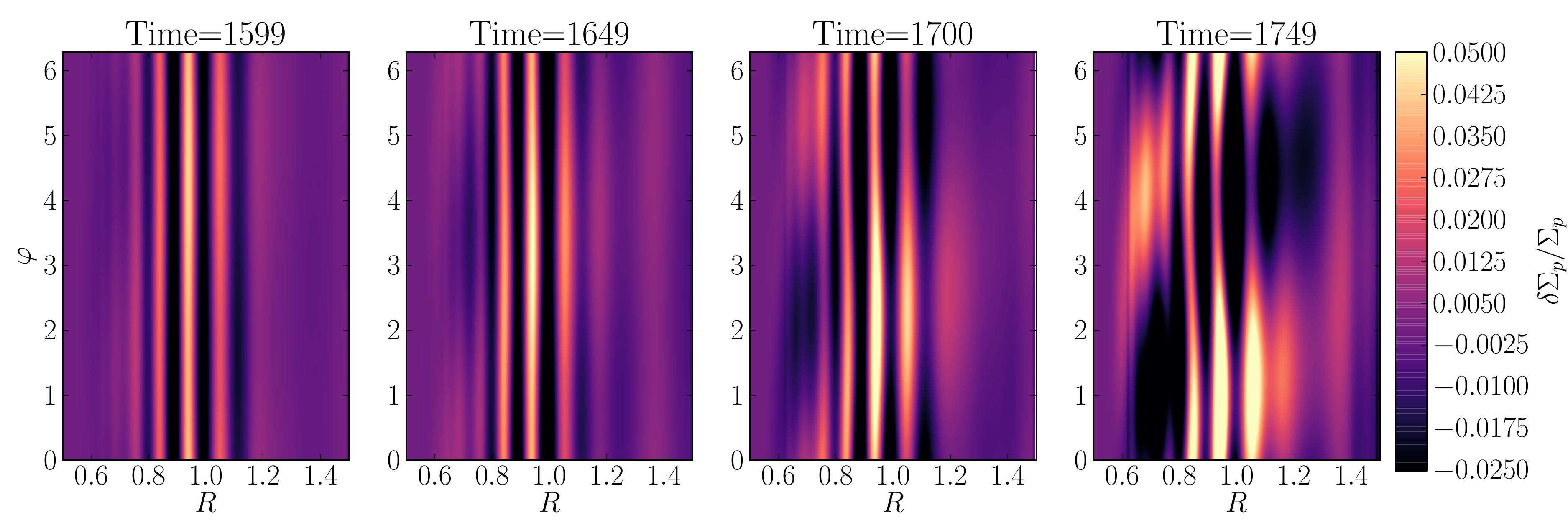}
\includegraphics[width=\textwidth]{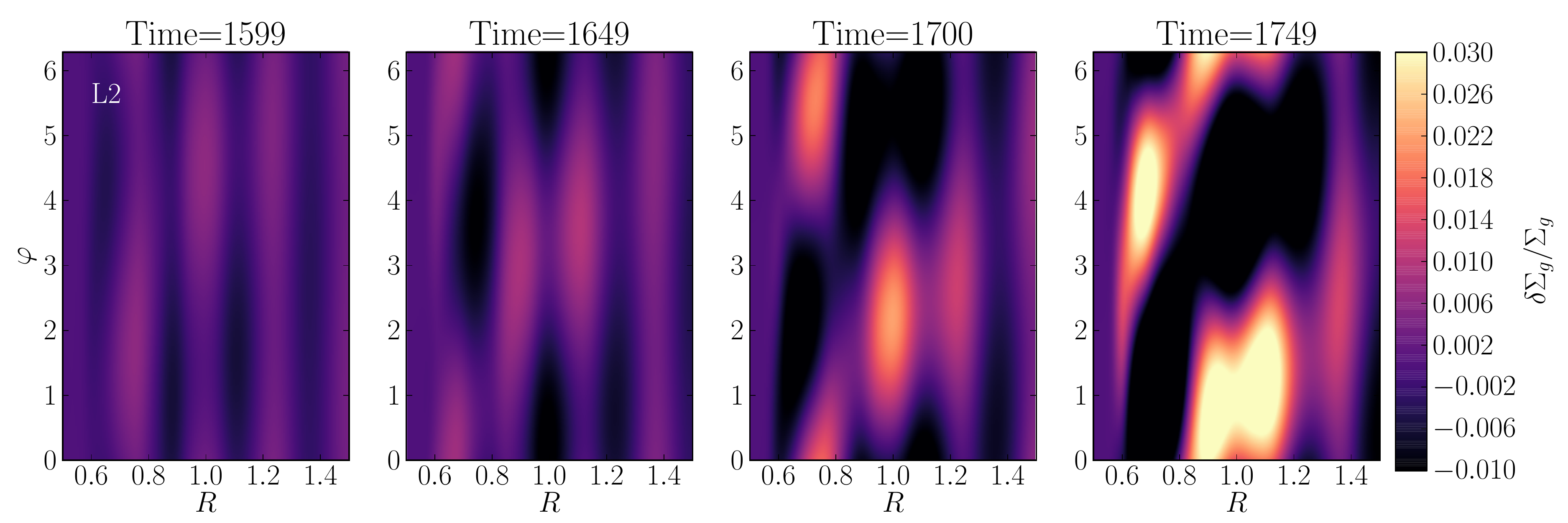}
\caption{{\it Top:} Relative dust density perturbation at different times for model L2 but in which  a diffusion pressure is employed for diffusion modelling.  {\it Bottom:} same but for the gas component.}
\label{fig:fig_modelea_2}
\end{figure*}

It has been shown by Tominaga et al. (2019) that  introducing a diffusion term as a source term in the continuity equation for the dust can violate the conservation of the total angular momentum of the dusty gas disc. To bypass this issue, Klahr \& Schreiber (2020) rather implemented the diffusion flux in the momentum equation for the dust rather than in the continuity equation. This can be done by including a "diffusion pressure" within the momentum equation for the dust which therefore becomes:

\begin{equation}
\frac{\partial {\bf V}}{\partial t}+({\bf V}\cdot \nabla){\bf V}=-\frac{\nabla P_p}{\Sigma_p}-\frac{D}{\tau_s}\frac{\nabla \Sigma_p}{\Sigma_p}-\frac{{\bf V}-{\bf v}}{\tau_s}-{\bf \nabla} \Phi_{\star}-{\bf \nabla} \Phi_{sg}-{\bf \nabla} \Phi_{ind}
\label{eq:dust}
\end{equation}

Here, we show that employing this alternative way to model dust diffusion lead to results very similar to those obtained using a more traditional diffusion term in the dust continuity equation. In particular, we check that the $m=1$ mode instability that we find for the smallest SGI growth rates is not related to unphysical angular momentum transport due to our modelling of dust diffusion. Fig. \ref{fig:linear} shows the temporal evolution of the relative perturbations of the dust and gas surface densities, and that of the dust-to-gas ratio. Again, the linear growth rate is found to be consistent with the analytical estimation from Tominaga et al. (2019), which is  a factor of $\sim 2$ larger than the one obtained by Takahashi et al. 2014 (see Fig. 2 in Tominaga et al. 2019). Comparing Fig. \ref{fig:linear_2} with Fig. \ref{fig:linear},  we see that there is no evidence for overstability in the case where diffusion is modelled using a diffusion pressure, which is also consistent with the results of Tominaga et al. (2019). Moreover, as revealed by Fig.  \ref{fig:fig_modelea_2} which presents contours of the dust and gas surface densities at different times,  $m=1$ spiral waves are found to develop when employing a diffusion pressure. Compared to model L2, the onset of the instability here occurs at earlier times, but this is likely to be related to a higher SGI linear growth rate. Nevertheless, this  demonstates that the diffusion modelling is irrelevant to the $m=1$ mode.

\end{document}